\documentclass[10pt,journal,compsoc]{IEEEtran}

\IEEEoverridecommandlockouts
\usepackage{amssymb}
\usepackage{algorithm}
\usepackage{algorithmic}
\usepackage{bibentry}
\usepackage{amsfonts}
\usepackage{subfigure}
\usepackage{multirow}
\usepackage{amsmath}
\usepackage{graphicx}
\usepackage{epsfig}
\usepackage{color}
\usepackage{epstopdf}
\usepackage{rotating}
\usepackage{caption}
\usepackage{float}
\usepackage{multicol}
\usepackage{amssymb}
\usepackage{graphicx}
\usepackage{bbding}
\usepackage{algorithm}
\usepackage{algorithmic}
\usepackage{balance}
\usepackage{microtype}
\usepackage{graphicx}
\usepackage{subfigure}
\usepackage{booktabs} 
\usepackage{multirow}
\usepackage{amsmath}
\usepackage{makecell}
\usepackage{mathtools}
\usepackage{etoolbox}
\usepackage{cases}
\usepackage{enumitem}
\usepackage[table,dvipsnames]{xcolor}
\usepackage{subfigure}

\usepackage{colortbl}
\usepackage{tabularx}
\usepackage{hhline}
\usepackage{threeparttable}
\usepackage{soul}
\usepackage{color}
\definecolor{brown}{RGB}{139,64,0}
\definecolor{pink}{RGB}{255,170,182}
\definecolor{purple}{RGB}{160,32,240}

\usepackage{amssymb}

\usepackage{cases}
 
\usepackage{cite}
\usepackage{amsmath,amssymb,amsfonts}
\usepackage{algorithmic}
\usepackage{graphicx}
\usepackage{textcomp}
\usepackage{xcolor}
\def\BibTeX{{\rm B\kern-.05em{\sc i\kern-.025em b}\kern-.08em
    T\kern-.1667em\lower.7ex\hbox{E}\kern-.125emX}}
    
\usepackage{mathtools}

\usepackage{amssymb}
\usepackage{amsthm}
\usepackage{color}
\usepackage{multirow}

\newcommand{\cut}[1]{{}}

\makeatletter
\let\@@span\span
\def\sp@n{\@@span\omit\advance\@multicnt\m@ne}
\makeatother

\newcommand{\bc}{\begin{center}}
\newcommand{\ec}{\end{center}}

\newcommand{\bdm}{\begin{displaymath}}
\newcommand{\edm}{\end{displaymath}}

\newcommand{\beq}{\begin{equation}}
\newcommand{\eeq}{\end{equation}}

\newcommand{\bfl}{\begin{flushleft}}
\newcommand{\efl}{\end{flushleft}}

\newcommand{\bt}{\begin{tabbing}}
\newcommand{\et}{\end{tabbing}}

\newcommand{\beqn}{\begin{align}}
\newcommand{\eeqn}{\end{align}}

\newcommand{\beqs}{\begin{align*}} 
\newcommand{\eeqs}{\end{align*}}  

\begin{document}
\title{Recommender Systems in the Era of\\ Large Language Models (LLMs)}







\author{Zihuai Zhao, Wenqi Fan, Jiatong Li, Yunqing Liu, Xiaowei Mei, Yiqi Wang, \\
Zhen Wen, Fei Wang, Xiangyu Zhao, Jiliang Tang, and Qing Li

\IEEEcompsocitemizethanks{
\IEEEcompsocthanksitem W. Fan is with the Department of Computing (COMP) and Department of Management and Marketing (MM), The Hong Kong Polytechnic University. E-mail: wenqifan03@gmail.com.
\IEEEcompsocthanksitem Z. Zhao, J. Li, Y. Liu, and Q. Li are with the Department of Computing, The Hong Kong Polytechnic University. E-mail: scofield.zzh@gmail.com, \{jiatong.li, yunqing617.liu\}@connect.polyu.hk, csqli@comp.polyu.edu.hk. 
\IEEEcompsocthanksitem  X. Mei is with the Department of Management and Marketing, The Hong Kong Polytechnic University. E-mail: michael.mei@polyu.edu.hk. 
\IEEEcompsocthanksitem Y. Wang is with National University of Defense Technology. E-mail: yiq@nudt.edu.cn.  This work was done when Yiqi Wang was a PhD student at Michigan State University.
\IEEEcompsocthanksitem Z. Wen and F. Wang are with Amazon. E-mail: \{zhenwen, feiww\}@amazon.com.
\IEEEcompsocthanksitem X. Zhao is with City University of Hong Kong. E-mail: xy.zhao@cityu.edu.hk.
\IEEEcompsocthanksitem J. Tang is with Michigan State University. E-mail: tangjili@msu.edu.

}
\thanks{(Corresponding authors: Wenqi Fan and Qing Li.)}
}

\markboth{IEEE TRANSACTIONS ON KNOWLEDGE AND DATA ENGINEERING, SUBMISSION 2023}%
{Shell \MakeLowercase{\textit{et al.}}: Bare Demo of IEEEtran.cls for Computer Society Journals}

\IEEEtitleabstractindextext{%
\begin{abstract}

With the prosperity of e-commerce and web applications, Recommender Systems (RecSys) have become an indispensable and important component in our daily lives, providing personalized suggestions that cater to user preferences.  While Deep Neural Networks (DNNs) have achieved significant advancements in enhancing recommender systems by modeling user-item interactions and incorporating their textual side information, these DNN-based methods still exhibit some limitations, such as difficulties in effectively understanding users' interests and capturing textual side information, inabilities in generalizing to various seen/unseen recommendation scenarios and reasoning on their predictions, etc.
Meanwhile, the development of Large Language Models (LLMs), such as ChatGPT and GPT-4,  has revolutionized the fields of Natural Language Processing (NLP) and Artificial Intelligence (AI), due to their remarkable abilities in fundamental responsibilities of language understanding and generation, as well as impressive generalization capabilities and reasoning skills. As a result, recent studies have actively attempted to harness the power of LLMs to enhance recommender systems. Given the rapid evolution of this research direction in recommender systems, there is a pressing need for a systematic overview that summarizes existing LLM-empowered recommender systems, so as to provide researchers and practitioners in relevant fields with an in-depth understanding. Therefore, in this survey, we conduct a comprehensive review of LLM-empowered recommender systems from various aspects including pre-training, fine-tuning, and prompting paradigms. More specifically, we first introduce the representative methods to harness the power of LLMs (as a feature encoder) for learning representations of users and items.  Then, we systematically review the emerging advanced techniques of LLMs for enhancing recommender systems from three paradigms, namely pre-training, fine-tuning, and prompting.  Finally, we comprehensively discuss the promising future directions in this emerging field.

\end{abstract}

\begin{IEEEkeywords}
 Recommender Systems,  Large Language Models (LLMs), Pre-training and Fine-tuning, Prompting, In-context Learning. 
\end{IEEEkeywords}}

\maketitle
\IEEEpeerreviewmaketitle

\section{Introduction}
\label{Introduction}

\IEEEPARstart{R}ecommender Systems (RecSys) play a vital role in alleviating information overload for enriching users' online experience (\textit{i.e.}, users need to filter overwhelming information to locate their interested information)~\cite{fan2020graph,chen2023fairly}. 
They offer personalized suggestions toward candidate items tailored to meet user preferences in various application domains, such as entertainment~\cite{gao2023chat}, e-commerce~\cite{chen2023knowledge}, and job matching~\cite{chen2023fairly}.
For example, in movie recommendations (\textit{e.g.}, \emph{IMDB} and \emph{Netflix}), the latest movies can be recommended to users based on the content of movies and the watch histories of users, which assists users in discovering new movies that accord with their interests.
The basic idea of recommender systems is to make use of the interactions between users and items and their associated side information, especially textual information like item descriptions, user profiles, and user reviews, to predict the matching score between users and items (\textit{i.e.}, the probability that the user would like the item)~\cite{fan2022comprehensive}. 
More specifically, collaborative behaviors between users and items have been leveraged to design various recommendation models, which can be further used to learn the representations of users and items~\cite{he2020lightgcn,fan2019deep_daso}.
In addition, textual side information of users and items contains rich knowledge that can assist in the calculation of the matching scores, which provides valuable insights into understanding user preferences for advancing recommender systems~\cite{zheng2017joint}.

Due to the remarkable ability of representation learning in various fields, Deep Neural Networks (DNNs) have been widely adopted to advance recommender systems~\cite{zhang2019deep,fan2023generative}. DNNs demonstrate distinctive abilities in modeling user-item interactions with different architectures. 
For example, as particularly effective tools for sequential data, Recurrent Neural Networks (RNNs) have been adopted to capture high-order dependencies in user interaction sequences~\cite{hidasi2015session,fan2019deep_dscf}.
Considering users' online behaviors (\textit{e.g.,} chick, purchase, socializing) as graph-structured data, Graph Neural Networks (GNNs) have emerged as advanced representation learning techniques to learn user and item representations~\cite{he2020lightgcn,fan2020graph,fan2019graph}.
Meanwhile, DNNs have also demonstrated advantages in encoding side information. 
For instance, a BERT-based method is proposed to extract and utilize textual reviews from users~\cite{qiu2021u}.

\begin{figure*}[t]
    \centering
    \includegraphics[width=0.95\textwidth]{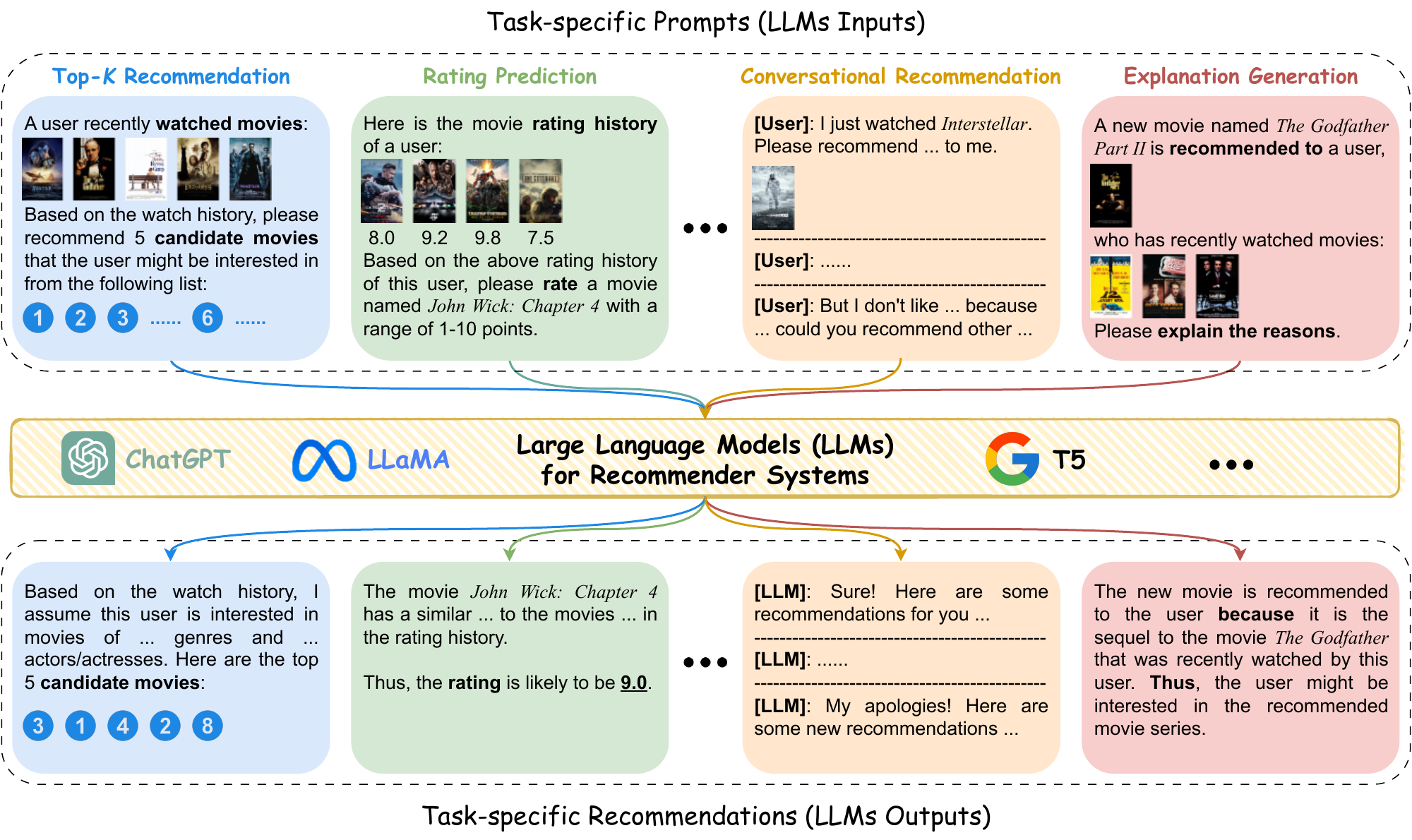}
    \caption{Examples of the application of LLMs for various recommendation tasks in the scenario of movie recommendations. In this workflow, LLMs directly act as recommenders through task-specific prompts, where textual information (or even multimodal data like images) are leveraged for recommendation tasks.}  
    \label{fig:LLMs tasks}  
\end{figure*}

Despite the aforementioned success, most existing advanced recommender systems still face some intrinsic limitations. 
\emph{First}, due to the limitations on model scale and data size, previous DNN-based models (\textit{e.g.}, CNN and LSTM) and pre-trained language models (\textit{e.g.}, BERT) for recommender systems cannot sufficiently capture textual knowledge about users and items, demonstrating their inferior natural language understanding capability, which leads to sub-optimal prediction performance in various recommendation scenarios. 
\emph{Second}, most existing RecSys methods have been specifically designed for their own tasks and have inadequate generalization ability to their unseen recommendation tasks.
For example, a recommendation algorithm is well-trained on a user-item rating matrix for predicting movies' rating scores, while it is challenging for this algorithm to perform top-$k$  movie recommendations along with certain explanations. 
This is due to the fact that the design of these recommendation architectures highly depends on task-specific data and domain knowledge toward specific recommendation scenarios such as top-$k$ recommendations, rating predictions, and explainable recommendations.
\emph{Third}, most existing DNN-based recommendation methods can achieve promising performance on recommendation tasks needing simple decisions (\textit{e.g.}, rating prediction, and top-$k$ recommendations). However, they face difficulties in supporting complex and multi-step decisions that involve multiple reasoning steps.  For instance, multi-step reasoning is crucial to trip planning recommendations, where RecSys should first consider popular tourist attractions based on the destination, then arrange a suitable itinerary corresponding to the tourist attractions, and finally recommend a journal plan according to specific user preferences (\textit{e.g.}, cost and time for travel).

Recently, as advanced natural language processing techniques, Large Language Models (LLMs) with billion parameters have generated large impacts on various research fields such as Natural Language Processing (NLP)~\cite{brown2020language}, Computer Vision~\cite{zhou2020unified}, and Molecule Discovery~\cite{li2023empowering}.
Technically, most existing LLMs are transformer-based models pre-trained on a vast amount of textual data from diverse sources, such as articles, books,  websites, and other publicly available written materials.
As the parameter size of LLMs continues to scale up with a larger training corpus, recent studies indicated that LLMs can lead to the emergence of remarkable capabilities~\cite{chen2023exploring,zhao2023survey}.
More specifically, LLMs have demonstrated the unprecedentedly powerful abilities of their fundamental responsibilities in language understanding and generation.
These improvements enable LLMs to better comprehend human intentions and generate language responses that are more human-like in nature.
Moreover, recent studies indicated that LLMs exhibit impressive generalization and reasoning capabilities, making LLMs better generalize to a variety of unseen tasks and domains.
To be specific, instead of requiring extensive fine-tuning on each specific task, LLMs can apply their learned knowledge and reasoning skills to fit new tasks simply by providing appropriate instructions or a few task demonstrations. 
Advanced techniques such as in-context learning can further enhance such generalization performance of LLMs without being fine-tuned on specific downstream tasks~\cite{zhao2023survey}. 
In addition, empowered by prompting strategies such as chain-of-thought, LLMs can generate the outputs with step-by-step reasoning in complicated decision-making processes. 
Hence, given their powerful abilities, LLMs demonstrate great potential to revolutionize recommender systems.

Very recently, initial efforts have been made to explore the potential of LLMs as a promising technique for the next-generation RecSys. 
For example, Chat-Rec~\cite{gao2023chat} is proposed to enhance the recommendation accuracy and explainability by leveraging ChatGPT to interact with users through conversations and then refine the candidate sets generated by traditional RecSys for movie recommendations.
Zhang et al.~\cite{zhang2023recommendation} employ T5 as LLM-based RecSys, which enables users to deliver their explicit preferences and intents in natural language as RecSys inputs, demonstrating better recommendation performance than merely based on user-item interactions. 
Figure~\ref{fig:LLMs tasks} demonstrates some examples of applying LLMs for various movie recommendation tasks, including top-$K$ recommendation, rating prediction, conversational recommendation, and explanation generation.
Due to their rapid evolution, it is imperative to comprehensively review recent advances and challenges of LLMs-empowered recommender systems.

Therefore, in this survey, we provide a systematic overview of LLM-empowered recommender systems in terms of \emph{pre-training},  \emph{fine-tuning}, and \emph{prompting} paradigms, which serve as three representative approaches to harness the power of LLMs ~\cite{min2023recent,zhao2023survey}. In particular, our survey is organized as follows.
First, we review the milestones in the field of RecSys and LLMs, respectively, and their combinations in Section~\ref{sec:related work}.
Then, two basic types of recommender systems that take advantage of LLMs to learn the representation of users and items are illustrated in Section~\ref{sec:representation}, namely the ID-based RecSys and the textual side information-enhanced RecSys.
Subsequently, we comprehensively summarize the advanced techniques for adapting LLMs to recommender systems in terms of pre-training \& fine-tuning and prompting paradigms in Section~\ref{sec:pre-train and fine-tune LLM4Rec} and Section~\ref{sec:prompting LLM4Rec}, respectively.
Finally, the emerging challenges posed by adapting LLMs to recommendations and some potential future directions are discussed in Section~\ref{sec:future_work}.

Recapping existing surveys in the domain of recommender systems, diverse focuses have been reviewed to facilitate the performance of RecSys from the perspective of deep learning techniques~\cite{gao2023survey,afsar2022reinforcement,wu2022graph,zhang2019deep}, evaluation methodology~\cite{alhijawi2022survey,zangerle2022evaluating}, trustworthiness~\cite{zehlike2022fairness,chen2023bias,wang2023survey} and other aspects. In the era of LLMs, the integration of LLMs into recommender systems has drawn increasing attention from recent studies,
which highlights the significance and necessity of systematically reviewing the emerging trends and advanced techniques in this interdisciplinary field of LLM-empowered recommender systems.
Before or concurrent to our survey, Liu \textit{et al.}~\cite{liu2023pre-recsys} review the training strategies and learning objectives of the language modeling paradigm adaptations for recommender systems. However, this work majorly examines early-stage language models for RecSys, such as BERT and GPT-2. Following the release of more advanced LLMs like ChatGPT and LLaMA, remarkable evolution has been brought to the adaption of LLMs in RecSys, which urges a more up-to-date review.
More recently, Wu \textit{et al.}~\cite{wu2023survey} summarize LLMs for recommender systems from discriminative and generative perspectives, which compares the two styles of LLMs tailored to their distinct abilities in recommendations. 
Meanwhile, Lin \textit{et al.}~\cite{lin2023can} introduce two orthogonal perspectives: where and how to adapt LLMs in recommender systems. In particular, this survey presents a pipeline of RecSys, reviewing the various functionalities of LLMs through the procedure of recommendations. 

Despite the aforementioned progress, existing surveys mainly emphasize the application aspects of LLMs in addressing their distinctive capabilities in RecSys, where the corresponding techniques proposed in the domain of LLMs are not systematically reviewed.
Therefore, our survey comprehensively reviews such domain-specific techniques for adapting LLMs to recommendations, which contributes to an in-depth understanding of developing LLM-based methods tailored to RecSys for future research.
\section{RELATED WORK}
\label{sec:related work}
\begin{figure*}[t]
    \centering
    \includegraphics[width=0.9\textwidth]{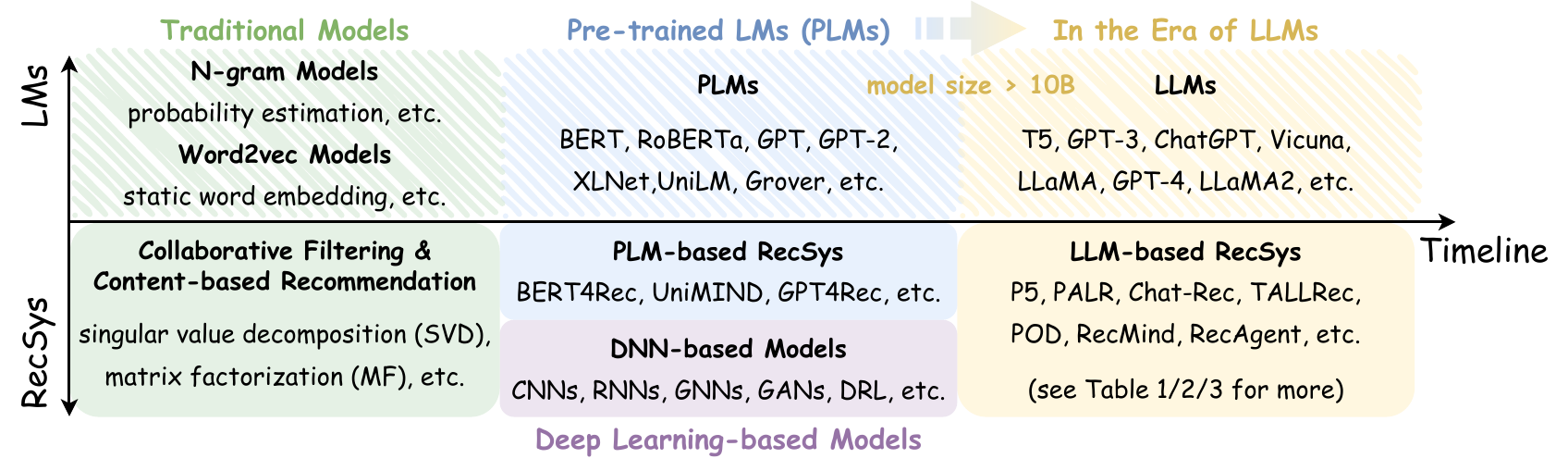}
    \caption{A timeline of milestones in the domains of recommender systems (RecSys) and language models (LMs). In order to align RecSys and LMs domains, the timeline is organized regardless of the exact time period but according to three stages: \emph{traditional models}, \emph{pre-trained language models}/\emph{deep learning-based models}, and \emph{the era of LLMs} as highlighted in colors.}
    \label{fig:milestones}  
\end{figure*}

In this section, we briefly review some related works on recommender systems, LLMs, and their combinations. As illustrated in Figure~\ref{fig:milestones}, a timeline of milestones in the domains of recommender systems and language models is provided, reviewing the development of the interdisciplinary field of LLM-empowered recommender systems.

\subsection{Recommender Systems (RecSys)}
To address the information overload problem, recommender systems have emerged as a crucial tool in various online applications by providing personalized content and services to individual users~\cite{wu2022disentangled,fan2022graph}. 
Typically, most existing recommendation approaches can fall into two main categories: Collaborative Filtering (CF) and  Content-based recommendation. 
As the most common technique, CF-based recommendation methods aim to find similar behavior patterns of users to predict the likelihood of future interactions~\cite{fan2019deep_dscf}, which can be achieved by utilizing the historical interaction behaviors between users and items, such as purchase history or rating data.  
For example, as one of the most popular CF methods,  Matrix Factorization (MF) is introduced to learn representations of users and items by using pure user-item interactions~\cite{fan2018deep,fan2019deep_daso}.
In other words, unique identities of users and items (\textit{i.e.}, discrete IDs) are encoded to continue embedding vectors so that the matching score can be calculated easily for recommendations~\cite{zhao2021autoloss,zhaok2021autoemb}.
Content-based recommendation methods generally take advantage of additional knowledge about users or items, such as user demographics or item descriptions, to enhance user and item representations for improving recommendation performance~\cite{vasile2016meta}. 
Note that as textual information is one of the most available contents for users and items, we mainly focus on text as content in this survey.  

Due to the remarkable representation learning capabilities,  deep learning techniques have been effectively applied to develop recommender systems~\cite{fan2022graph,fan2022comprehensive}.
For instance, NeuMF is proposed to model non-linear interactions between users and items by replacing the general inner product with DNNs~\cite{he2017neural}.  
Considering that data in RecSys can be naturally represented as graph-structured data, GNN techniques are treated as the main deep learning approaches for learning meaningful representations of nodes (\textit{i.e.}, users and items) via message propagation strategies for recommender systems~\cite{ying2018graph,fan2020graph,ma2021deep,derr2020epidemic}. 
In order to integrate textual knowledge about users and items, DeepCoNN is developed to use CNNs to encode users' reviews written for items with two parallel neural networks so as to contribute to rating predictions in recommender systems~\cite{zheng2017joint}.
Meanwhile, a neural attention framework NARRE is introduced to simultaneously predict users' ratings towards items and generate review-level explanations for the predictions~\cite{chen2018neural}.

Recently, language models have been increasingly utilized in recommender systems due to their capacity to comprehend and produce human natural language. 
These models are designed to comprehend the semantics and syntax of human natural language, thereby enabling RecSys to provide more personalized recommendations, such as news recommendations~\cite{wu2020mind,wu2023personalized}, and drug recommendations~\cite{dongre2023deep}.
Specifically, a sequential recommendation method called BERT4Rec is proposed to adopt  Bidirectional Encoder Representations from Transformers (\textit{i.e.}, BERT) to model the sequential nature of user behaviors~\cite{sun2019bert4rec}. 
Furthermore, to take advantage of Transformer's capability for language generation, Li \textit{et al.}~\cite{liu2023chatgpt} design a transformer-based framework to simultaneously make item recommendations and generate explanations in recommender systems.

\subsection{From Pre-trained Language Models (PLMs) to Large Language Models (LLMs)}

As a type of advanced Artificial Intelligence (AI) techniques, LLMs are trained on a large amount of textural data with billions of parameters to understand the patterns and structures of natural language. 
There are several classical types of pre-trained language models (PLMs) available, such as BERT (Bidirectional Encoder Representations from Transformers)~\cite{devlin2018bert}, GPT (Generative Pre-trained Transformer)~\cite{radford2018improving}, and T5 (Text-To-Text Transfer Transformer)~\cite{raffel2020exploring}.
Typically, these language models fall into three main categories: encoder-only models, decoder-only models, and encoder-decoder models. 

BERT, GPT, and T5 are distinct models based on the Transformer architecture~\cite{vaswani2017attention}. More specifically, BERT, an encoder-only model, uses bi-directional attention to process token sequences, considering both the left and right context of each token. It is pre-trained based on massive amounts of text data using tasks like masked language modeling and next-sentence prediction, thereby capturing the nuances of language and meaning in context. 
This process translates text into a vector space, facilitating nuanced and context-aware analyses. 
On the other hand, GPT, based on the transformer decoder architecture, uses a self-attention mechanism for one-directional word sequence processing from left to right. GPT is mainly adopted in language generation tasks, mapping embedding vectors back to text space, and generating contextually relevant responses. 
At last, T5, an encoder-decoder model, could handle any text-to-text task by converting every natural language processing problem into a text generation problem. For instance, it can re-frame a sentiment analysis task into a text sequence, like '\emph{sentiment: I love this movie.}', which adds '\emph{sentiment:}' before '\emph{I love this movie.}'. Then it will get the answer '\emph{positive}'. 
By doing so, T5 uses the same model, objective, and training procedure for all tasks, making it a versatile tool for various NLP tasks.

Due to the increasing scale of models, LLMs have revolutionized the field of NLP by demonstrating unprecedented capabilities in understanding and generating human-like textual knowledge~\cite{chen2023exploring,zhang2023certified}.
These models (e.g., GPT-3~\cite{brown2020language}, LaMDA~\cite{thoppilan2022lamda}, PaLM~\cite{chowdhery2022palm}, and Vicuna~\cite{chiang2023vicuna}) often based on transformer architectures, undergo training on extensive volumes of text data. 
This process enables them to capture complex patterns and nuances in human language. 
Recently, LLMs have demonstrated remarkable capabilities of ICL, a concept that is central to their design and functionality. 
ICL refers to the model's capacity to comprehend and provide answers based on the input context as opposed to merely relying on inside knowledge obtained through pre-training.
Several works have explored the utilization of ICL in various tasks, such as SG-ICL~\cite{kim2022self} and EPR~\cite{rubin2021learning}. 
These works show that ICL allows LLMs to adapt their responses based on input context instead of generating generic responses.
Another technique that can enhance the reasoning abilities of LLMs is chain-of-thought (CoT). This method involves supplying multiple demonstrations to describe the chain of thought as examples within the prompt, guiding the model's reasoning process~\cite{wei2022chain}. 
An extension of the CoT is the concept of self-consistency, which operates by implementing a majority voting mechanism on answers~\cite{wang2022self}. 
Current researches continue to delve into the application of CoT in LLMs, such as STaR~\cite{zelikman2022star}, THOR~\cite{fei2023reasoning}, and Tab-CoT~\cite{jin2023tab}.
By offering a set of prompts to direct the model's thought process, CoT enables the model to reason more effectively and deliver more accurate responses. 

With the powerful abilities mentioned above, LLMs have shown remarkable potential in various fields, such as chemistry~\cite{li2023empowering}, education~\cite{kasneci2023chatgpt}, and finance~\cite{wu2023bloomberggpt}. 
These models, such as ChatGPT, have also been instrumental in enhancing the functionality and user experience of RecSys. 
One of the key applications of LLMs in RecSys is the prediction of user ratings for items. This is achieved by analyzing historical user interactions and preferences, which in turn enhances the accuracy of the recommendations~\cite{kang2023llms,zhiyuli2023bookgpt}. 
LLMs have also been employed in sequential recommendations, which analyze the sequence of user interactions to predict their next preference, such as TALLRec~\cite{bao2023tallrec}, M6-Rec~\cite{cui2022m6}, PALR~\cite{chen2023palr}, and P5~\cite{geng2022recommendation}.
Moreover, LLMs, particularly ChatGPT, have been utilized to generate explainable recommendations. One such example is Chat-Rec~\cite{gao2023chat}, which leverages ChatGPT to provide clear and comprehensible reasoning behind its suggestions, thereby fostering trust and user engagement. 
Furthermore, the interactive and conversational capabilities of LLMs have been harnessed to create a more dynamic recommendation experience.
For instance,  UniCRS~\cite{wang2022towards} develops a knowledge-enhanced prompt learning framework to fulfill both conversation and recommendation subtasks based on a pre-trained language model.
UniMIND~\cite{deng2023unified} proposes a unified multi-task learning framework by using prompt-based learning strategies in conversational recommender systems. 
Furthermore, it is worth noting that to investigate the potential of LLMs in learning on graphs, Chen \textit{et al.}~\cite{chen2023exploring} introduce two possible pipelines: \emph{LLMs-as-Enhancers} (\textit{e.g.}, LLMs enhance the textual information of node attributes) and \emph{LLMs-as-Predictors} (\textit{e.g.}, LLMs serve as independent predictor in graph learning like link prediction problems), which provide guidance on the design of LLMs for graph-based recommendations.

\section{Deep Representation Learning for LLM-based Recommender Systems}
\label{sec:representation}

\begin{figure*}[t]
    \centering
    \includegraphics[width=0.95\textwidth]{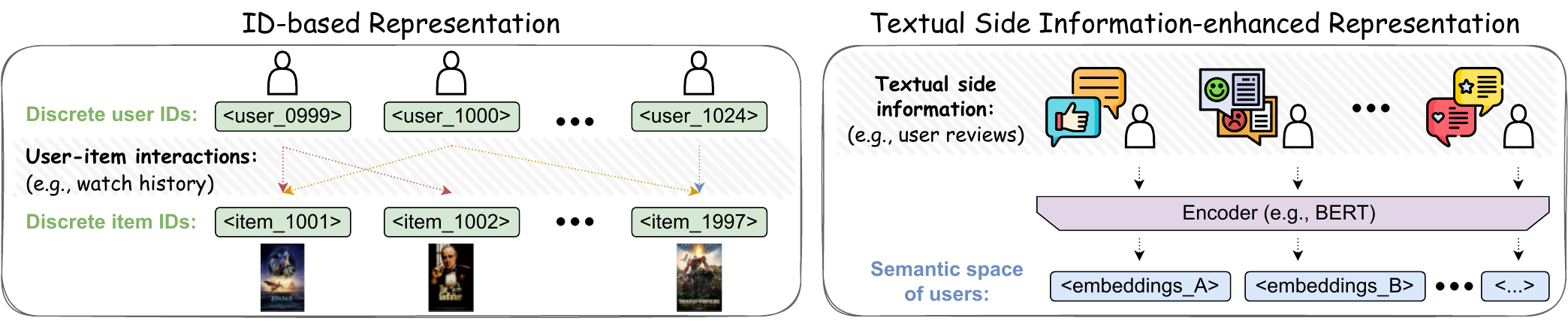}
    \caption{An illustration of two methods for representing users and items in LLM-based recommender systems: \emph{ID-based Representation} which denotes user-item interactions with discrete identities, and \emph{Textual Side Information-enhanced Representation} which leverages textual side information of users and items, such as user reviews of items.} 
    \label{fig:representation}  
\end{figure*}

Users and items are atomic units of recommender systems. 
To denote items and users in recommender systems, the straightforward method assigns each item or user a unique index (\textit{i.e.}, discrete IDs).
To capture users' preferences towards items, ID-based recommender systems are proposed to learn representations of users and items from user-item interactions. 
In addition, since textual side information about users and items provides rich knowledge to understand users' interests, textual side information-enhanced recommendation methods are developed to enhance user and item representation learning in an end-to-end training manner for recommender systems. In this section, we will introduce these two categories that take advantage of language models in recommender systems. 
These two kinds of recommender systems are illustrated in Figure \ref{fig:representation}.

\subsection{ID-based Recommender Systems}
Recommender systems are commonly used to affect users' behaviors for making decisions from a range of candidate items. 
These user behaviors (\textit{e.g.}, click, like, and subscription) are generally represented as user-item interactions, where users and items are denoted as discrete IDs.
Modern recommendation approaches are proposed to model these behaviors by learning embedding vectors of each ID representation. 
Generally, in LLM-based recommendation systems, an item or a user can be represented by a short phrase in the format of $``[prefix]\_[ID]"$, where the prefix denotes its type (\textit{i.e.}, item or user) and the ID number helps identify its uniqueness.

As the early exploration of LLM-based methods,  a unified paradigm called P5 is proposed to facilitate the transfer of various recommendation data formats~\cite{geng2022recommendation}, such as user-item interactions, user profiles, item descriptions, and user reviews, into natural language sequences by mapping users and items into indexes.
Note that the pre-trained T5 backbone is used to train the P5 with personalized prompts. 
Meanwhile, P5 incorporates the normal index phrase with a pair of angle brackets to treat these indexes as special tokens in the vocabulary of LLMs (e.g., $<item\_6637>$), avoiding tokenizing the phrases into separate tokens.

Based on P5, Hua et al. put forward four straightforward but effective indexing solutions~\cite{hua2023index}: sequential indexing, collaborative indexing, semantic (content-based) indexing, and hybrid indexing, underscoring the significance of indexing methods.
Different from P5's randomly assigning numerical IDs to each user or item, Semantic IDs, a tuple of codewords with semantic meanings for each user or item, is proposed to serve as unique identifiers, each carrying semantic meaning for a particular user or item ~\cite{rajput2023recommender}. 
Meanwhile, to generate these codewords, a hierarchical method called RQ-VAE is also proposed ~\cite{rajput2023recommender} to leverage Semantic IDs, where recommendation data formats can be effectively transformed into natural language sequences for transformer-based models.

\subsection{Textual Side Information-enhanced Recommender Systems}
Despite the aforementioned success,  ID-based methods suffer from intrinsic limitations. 
That is due to the fact that pure ID indexing of users and items is naturally discrete, which cannot provide sufficient semantic information to capture representations of users and items for recommendations. 
As a result, it is very challenging to perform relevance calculations based on index representations among users and items, especially when user-item interactions are severely sparse. 
Meanwhile, ID indexing usually requires modifying the vocabularies and altering the parameters of LLMs, which brings additional computation costs. 

To address these limitations, a promising alternative solution is to leverage textual side information of users and items, 
which includes user profiles, user reviews for items, and item titles or descriptions. Specifically, given the textual side information of an item or a user, language models like BERT can serve as the text encoder to map the item or user into the semantic space, where we can group similar items or users and figure out their differences in a more fine-grained granularity. For instance, Li \textit{et al.} have investigated the performance comparison between ID and modality-based recommender systems, showing that ID-based recommender systems might be challenged by recommender systems that can better utilize side information~\cite{zheng2023where}.
Meanwhile, Unisec~\cite{hou2022towards} is one such approach that takes advantage of item descriptions to learn transferable representations from various recommendation scenarios. More specifically, Unisec also introduces a lightweight item encoder to encode universal item representations by using parametric whitening and a mixture-of-experts (MoE) enhanced adaptor. 
Besides, text-based collaborative filtering (TCF) is also explored by prompting LLMs like GPT-3~\cite{Li2023tcf}. Compared to the previous ID-based collaborative filtering, TCF methods demonstrate positive performance, proving the potential of textual side information-enhanced recommender systems. 

However, solely relying on language models to encode item descriptions might excessively emphasize text features. 
To mitigate this issue, VQ-Rec~\cite{hou2023learning} proposes to learn vector-quantized item representations, which can map item text into a vector of discrete indices (\textit{i.e.}, item codes) and use them to retrieve item representations from a code embedding table in recommendations. 
Beyond text features, Fan \textit{et al.}~\cite{fan2023zero} propose a novel method for the Zero-Shot Item-based Recommendation (ZSIR), focusing on introducing a Product Knowledge Graph (PKG) to LLMs to refine item features. More specifically, 
user and item embeddings are learned via multiple pre-training tasks upon the PKG.
Moreover, ShopperBERT~\cite{shin2021one4all} 
investigates modeling user behaviors to denote user representations in e-commerce recommender systems, which pre-trains user embedding through several pre-training tasks based on user purchase history.
Furthermore, IDA-SR~\cite{shin2021one4all}, an ID-Agnostic User Behavior Pre-training framework for Sequential Recommendation, directly retains representations from text information using pre-trained language models like BERT.
Specifically, given an item $i$ and its description with $m$ tokens $D_i = \{t_1, t_2, ..., t_m\}$, an extra start-of-sequence token $[CLS]$ is added to the description $D_i = \{[CLS], t_1, t_2, ..., t_m\}$. Then, the description is fed as the input to LLMs. Finally, the embedding of the token $[CLS]$ could be used as the ID-agnostic item representation.

\begin{figure*}[htbp]
    \centering
    \includegraphics[width=0.95\textwidth]{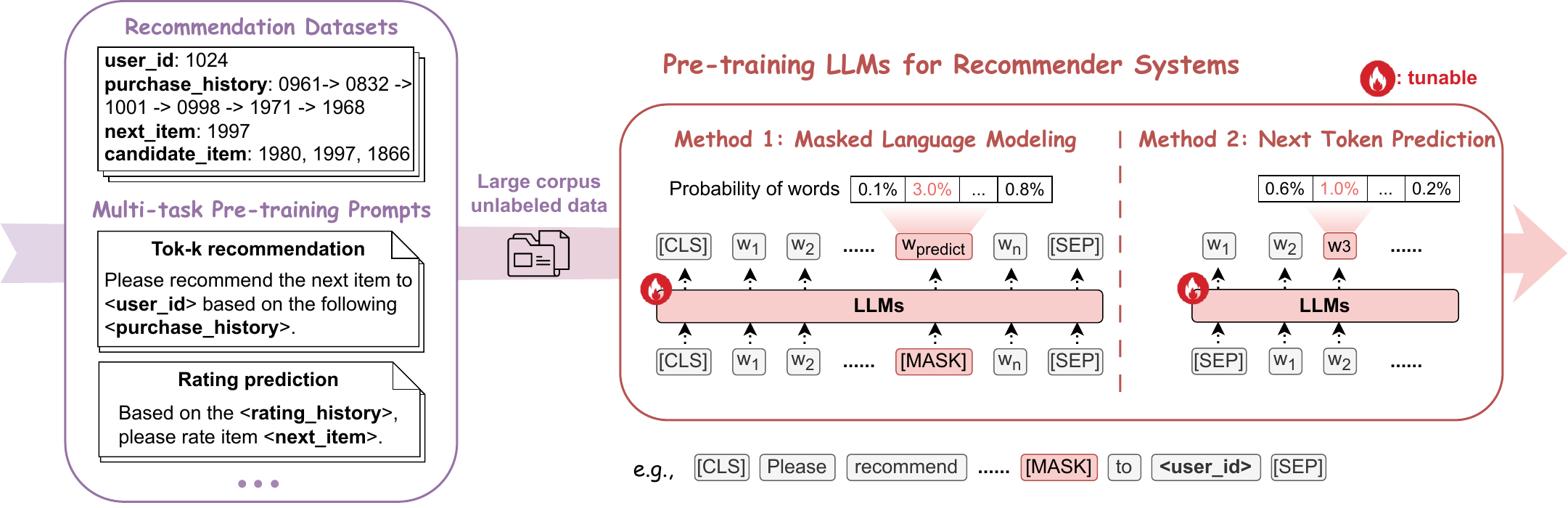}
    \caption{A workflow of pre-training LLMs for recommender systems in terms of two representative methods: \emph{Masked Language Modeling} which randomly masks tokens or spans in the sequence and requires LLMs to generate the masked tokens or spans based on the remaining context, and \emph{Next Token Prediction} which requires prediction for the next token based on the given context.} 
    \label{fig:pre-training}  
\end{figure*}

\begin{table*}[htbp]
\centering
\caption{Pre-training methods for LLM-empowered RecSys.}
\label{tab:pre-train}
\resizebox{1.5\columnwidth}{!}{
\begin{tabular}{c|c|c|c}
\toprule
Paradigms                     & Methods                                                       & Pre-training Tasks         & Code Availability                               \\ \midrule
\multirow{4}{*}{Pre-training} & \multirow{2}{*}{PTUM~\cite{wu2020ptum}} & Masked Behavior Prediction & \multirow{2}{*}{\url{https://github.com/wuch15/PTUM}} \\ \cline{3-3}
                              &                                                               & Next K Behavior Prediction &                                                 \\ \cline{2-4} 
                              & M6~\cite{cui2022m6}                     & Auto-regressive Generation & Not available                                   \\ \cline{2-4} 
                              & P5~\cite{geng2022recommendation}        & Multi-task Modeling        & \url{https://github.com/jeykigung/P5}                 \\ \bottomrule
\end{tabular}
}
\end{table*}

\section{Pre-training \& Fine-tuning LLMs for Recommender Systems} 
\label{sec:pre-train and fine-tune LLM4Rec}

\begin{figure*}[h]
    \centering
    \includegraphics[width=0.95\textwidth]{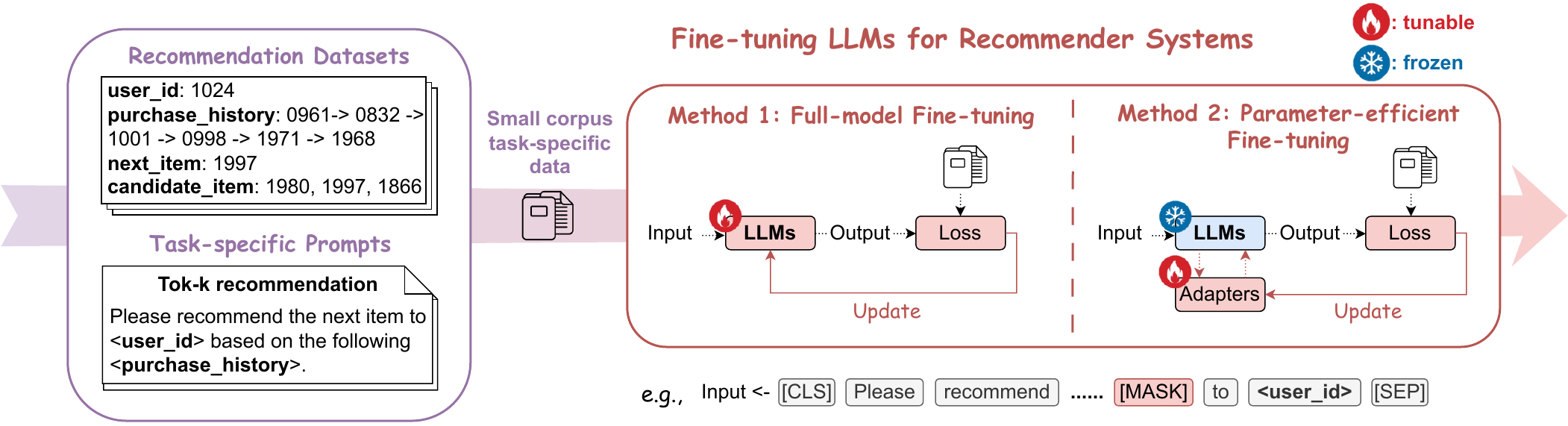}
    \caption{A workflow of fine-tuning LLMs for recommender systems in terms of two representative methods: \emph{Full-model Fine-tuning} which involves changing the entire model weights, and \emph{Parameter-efficient Fine-tuning} which involves fine-tuning a small proportion of model weights or a few extra trainable weights while fixing most of the parameters in LLMs.} 
    \label{fig:fine-tuning}  
\end{figure*}

In general, there are three key manners in developing and deploying LLMs in recommendation tasks, namely, \emph{pre-training, fine-tuning}, and \emph{prompting}. In this section, we first introduce the pre-training and fine-tuning paradigms, which are shown in Figure ~\ref{fig:pre-training} and Figure ~\ref{fig:fine-tuning}, respectively. More specifically, we will focus on the specific pre-training tasks applied in LLMs for recommender systems and fine-tuning strategies for better performance in downstream recommendation tasks. 
Note that the works mentioned below are summarized in Table \ref{tab:pre-train} and Table \ref{tab:fine-tune}.

\begin{table*}[htb]
\centering
\caption{Fine-tuning methods for LLM-empowered RecSys.}
\label{tab:fine-tune}
\resizebox{1.9\columnwidth}{!}{
\begin{tabular}{lcc}
\toprule
\multicolumn{1}{c|}{Paradigms}                    & \multicolumn{1}{c|}{Methods}                         & References \\ \midrule
\multicolumn{1}{c|}{\multirow{2}{*}{Fine-tuning}} & \multicolumn{1}{c|}{Full-model Fine-tuning}          & ~\cite{friedman2023leveraging}, ~\cite{shen2023towards}, ~\cite{wang2022transrec},~\cite{carranza2023privacy},~\cite{zheng2023generative},~\cite{kim2021intent}, and ~\cite{mao2023unitrec}$^1$        \\ \cline{2-3} 
\multicolumn{1}{c|}{}                             & \multicolumn{1}{c|}{Parameter-efficient Fine-tuning} & ~\cite{bao2023tallrec}$^2$, ~\cite{wu2023exploring}, and ~\cite{cui2022m6}          \\ \midrule
\multicolumn{3}{l}{\footnotesize Code Availability: \footnotesize$^1$\url{https://github.com/veason-silverbullet/unitrec}, \footnotesize$^2$\url{https://github.com/sai990323/tallrec}}                                                                                \\
\bottomrule
\end{tabular}
}
\end{table*}

\subsection{Pre-training Paradigm for Recommender Systems}
Pre-training is an important step in developing LLMs, which inherits the idea of transfer learning.
It involves training LLMs on a vast amount of corpus consisting of diverse and unlabeled text data. 
This strategy enables LLMs to acquire a broad understanding of various linguistic aspects, including grammar, syntax, semantics, and even common sense reasoning.
Through pre-training, LLMs can learn to recognize and generate coherent and contextually appropriate responses. 
In general, there are two mainstream paradigms to pre-train LLMs from the view of Natural Language Processing, while the selection of the pre-training strategy depends on the specific model structure.
For encoder-only or encoder-decoder Transformer structures, \emph{Masked Language Modeling} (MLM) is widely adopted, which randomly masks tokens or spans in the sequence and requires LLMs to generate the masked tokens or spans based on the remaining context ~\cite{kenton2019bert}. 
At the same time, \emph{Next Token Prediction} (NTP) is deployed for pre-training decoder-only Transformer structures, which requires prediction for the next token based on the given context ~\cite{radford2018improving}. 
Both the two pre-training tasks involve completing conditional sentences, but there are differences in their approaches. The Masked Language Model (MLM) task predicts masked tokens in a bi-directional context, while the Next Sentence Prediction (NTP) task only considers the previous context. As a result, MLM could assist LLMs in better understanding the meanings of tokens, while NTP is more natural for language generation tasks.
 
In recommender systems, most of the existing works follow the two classical pre-training paradigms. Next, we will introduce several representative methods. 
PTUM~\cite{wu2020ptum} proposes two similar pre-training tasks, Masked Behavior Prediction (MBP) and Next K behavior Prediction (NBP), to model user behaviors in recommender systems. 
Unlike language tokens, user behaviors are more diverse and thus more difficult to predict. 
In this case, instead of masking a span of tokens, PTUM only masks a single user behavior with the goal of predicting the masked behavior based on the other behaviors in the interaction sequence of the target user. 
On the other side, NBP models the relevance between past and future behaviors, which is crucial for user modeling. The goal of NBP is to predict the next $k$ behaviors based on the user-item interaction history. Considering the time sequence of user behaviors, NBP could naturally simulate the users and thus demonstrate better performance.

M6~\cite{cui2022m6} also adopts two pre-training objectives motivated by the two classical pre-training tasks, namely a text-infilling objective and an auto-regressive language generation objective, corresponding to the above two pre-training tasks, respectively. 
To be more specific, the text-infilling objective exhibits the pre-training task of BART~\cite{lewis2020bart}, which randomly masks a span with several tokens in the text sequence and predicts these masked spans as the pre-training target, providing the capability to assess the plausibility of a text or an event in the recommendation scoring tasks. 
Meanwhile, the auto-regressive language generation objective follows the Next Token Prediction task in natural language pre-training, but it is slightly different as it predicts the unmasked sentence based on the masked sequence.

Additionally, P5 adopts multi-mask modeling and mixes datasets of various recommendation tasks for pre-training. In this case, it can be generalized to various recommendation tasks and even unseen tasks with zero-shot generation ability ~\cite{geng2022recommendation}. Across different recommendation tasks, P5 applies a unified indexing method for representing users and items in language sequence as stated in Section \ref{sec:representation} so that the Masked Language Modelling task could be employed.

\subsection{Fine-tuning Paradigm for Recommender Systems}\label{sec:fine-tuning}
Fine-tuning is a crucial step in deploying pre-trained LLMs for specific downstream tasks. Especially for recommendation tasks, LLMs require fine-tuning to grasp more domain knowledge. 
Particularly, the fine-tuning paradigm involves training the pre-trained model based on task-specific recommendation datasets that include user-item interaction behaviors (\textit{e.g.}, purchase, click, ratings) and side knowledge about users and items (\textit{e.g.}, users' social relations and items' descriptions). 
This process allows the model to specialize its knowledge and parameters to improve performance in the recommendation domain. 
In general, fine-tuning strategies can be divided into two categories according to the proportion of model weights changed to fit the given task. 
One is \emph{full-model fine-tuning}, which changes the entire model weights in the fine-tuning process. 
By considering the computation cost, the other is \emph{parameter-efficient fine-tuning}, which aims to change only a small part of weights or develop trainable adapters to fit specific tasks.

\subsubsection{Full-model Fine-tuning}
As a straightforward strategy in deploying pre-trained LLMs to fit specific downstream recommendation tasks, full-model fine-tuning involves changing the entire model weights.
For example, RecLLM~\cite{friedman2023leveraging} is proposed to fine-tune LaMDA as a Conversational Recommender System (CRS) for YouTube video recommendation. Meanwhile, GIRL~\cite{zheng2023generative} leverages a supervised fine-tuning strategy for instructing LLMs in job recommendation. 
However, directly fine-tuning LLMs might bring unintended bias into recommender systems, producing serious harm toward specific groups or individuals based on sensitive attributes such as gender, race, and occupation.
To mitigate such harmful effects, a simple LLMs-driven recommendation (LMRec)~\cite{shen2023towards} is developed to alleviate the observed biases through train-side masking and test-side neutralization of non-preferential entities, which achieves satisfying results without significant performance drops. 
TransRec~\cite{wang2022transrec} studies pre-trained recommender systems in an end-to-end manner, by directly learning from the raw features of the mixture-of-modality items  (\textit{i.e.}, texts and images).
In this case, without relying on overlapped users or items, TransRec can be effectively transferred to different scenarios.
Additionally, Carranza \textit{et al.}~\cite{carranza2023privacy} propose privacy-preserving large-scale recommender systems by applying differentially private (DP) LLMs, which relieves certain challenges and limitations in DP training.

Contrastive learning has also emerged as a popular approach for fine-tuning LLMs in recommender systems. Several methods have been proposed in this direction.
SBERT~\cite{kim2021intent} introduces a triple loss function, where an intent sentence is paired with an anchor, and corresponding products are used as positive and negative examples in the e-commerce domain. 
Additionally, UniTRec~\cite{mao2023unitrec} proposes a unified framework that combines discriminative matching scores and candidate text perplexity as contrastive objectives to improve text-based recommendations. 

\subsubsection{Parameter-efficient Fine-tuning}
Full-model fine-tuning requires large computational resources as the size of LLMs scales up. Currently, it is infeasible for a single consumption-level GPU to fine-tune the most advanced LLMs, which usually have more than 10 billion parameters. In this case, Parameter-efficient Fine-tuning (PEFT) targets fine-tuning LLMs efficiently with lower requirements for computational resources. PEFT involves fine-tuning a small proportion of model weights or a few extra trainable weights while fixing most of the parameters in LLMs to achieve comparable performance with full-model fine-tuning. 

Currently, the most popular PEFT methods lie in introducing extra trainable weights as adapters. The adapter structure is designed for embedding into the transformer structure of LLMs~\cite{houlsby2019parameter}. 
For each Transformer layer, the adapter module is added twice: the first module is added after the projection following the multi-head attention, and the other is added after the two feed-forward layers. During fine-tuning, the original weights of pre-trained LLMs are fixed, while the adapters and layer normalization layers are fine-tuned to fit downstream tasks.
Thus, adapters contribute to the expansion and generalization of LLMs, relieving the problem of full-model fine-tuning and catastrophic forgetting.
Inspired by the idea of adapters and low intrinsic ranks of weight matrices in LLMs, Low-Rank Adaptation of LLMs (LoRA)~\cite{hu2021lora} introduces low-rank decomposition to simulate the change of parameters. Basically, LoRA adds a new pathway to specific modules handling matrix multiplication in the original structure of the LLMs. In the pathway, two serial matrices first reduce the dimension to a pre-defined dimension of the middle layer and then increase the dimension back. In this case, the dimension of the middle layer could simulate the intrinsic rank.

In recommender systems, PEFT can greatly reduce the computational cost of fine-tuning LLMs for recommendation tasks, which requires less update and maintains most of the model capabilities. TallRec~\cite{bao2023tallrec} introduces an efficient and effective tuning framework on the LLaMA-7B model and LoRA for aligning LLMs with recommendation tasks, which can be executed on a single RTX 3090. GLRec~\cite{wu2023exploring} takes advantage of LoRA for fine-tuning and adapting LLMs as job recommenders. LLaRA \cite{liao2023llara} also utilizes LoRA for fine-tuning LLMs, enabling LLMs to fit different tasks.
Moreover, M6~\cite{cui2022m6} applies LoRA fine-tuning, making it feasible to deploy LLMs in phone devices.

\section{Prompting LLMs for Recommender Systems} \label{sec:prompting LLM4Rec}

\begin{figure*}[t]
    \centering
    \includegraphics[width=0.9\textwidth]{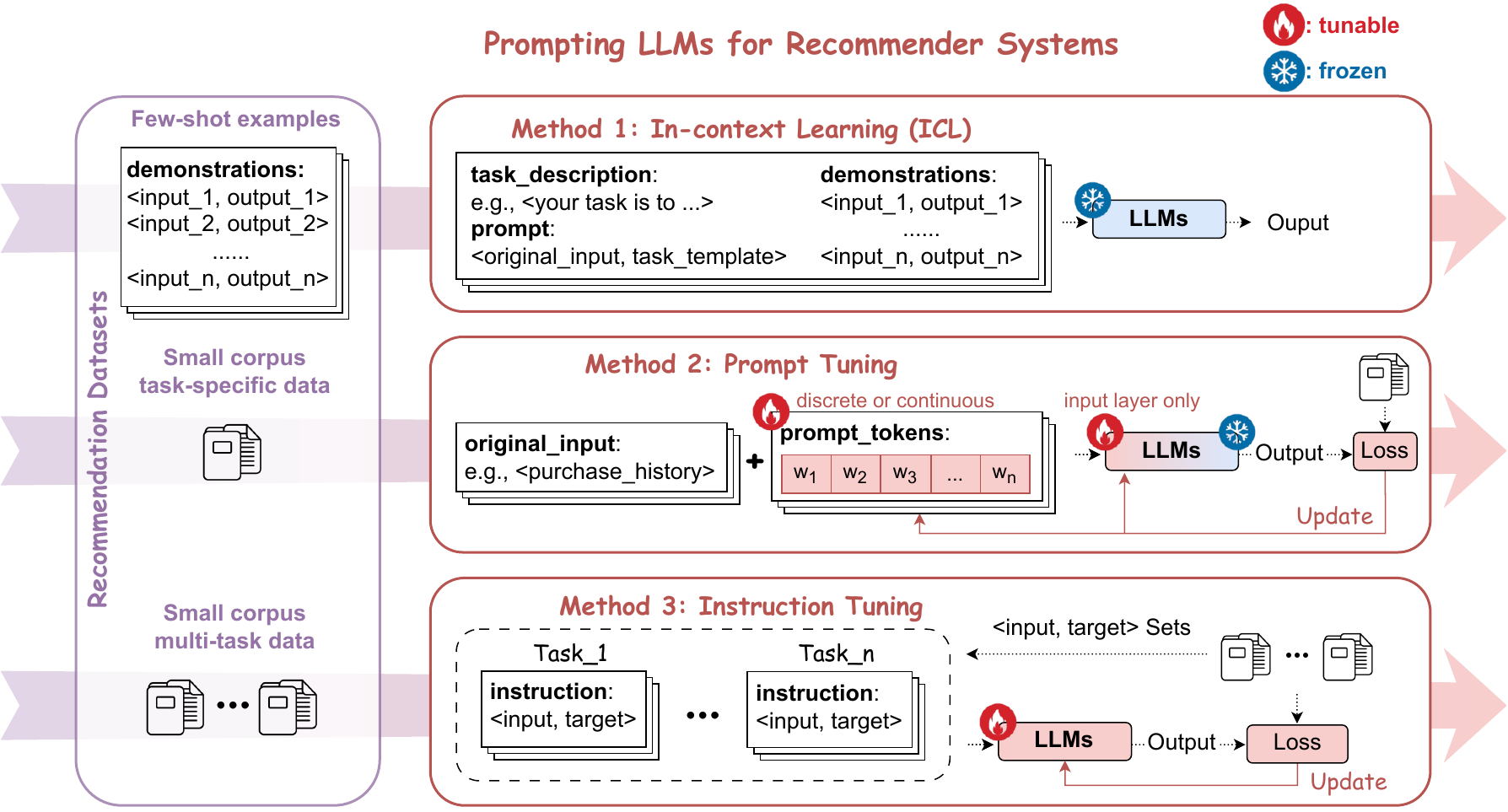}
    \caption{A workflow of prompting LLMs for recommender systems in terms of three representative methods: \emph{In-context Learning} (top) which requires no parameter update of LLMs, \emph{Prompt Tuning} (middle) which adds new prompt tokens to LLMs and optimizes the prompt along with minimal parameter updates at the input layer of LLMs, and \emph{Instruction Tuning} (bottom) which fine-tunes LLMs over multiple tasks-specific prompts, also known as instructions.} 
    \label{fig:prompting}  
\end{figure*}

\begin{table*}[!t] 
\centering
\caption{Prompting, prompt tuning, and instruction tuning methods for LLM-empowered RecSys. In particular, we categorize the integration of LLMs and RecSys into three representative approaches: \colorbox{cyan!20}{$\langle$ \emph{LLMs act as recommender} $\rangle$} (\textit{e.g.}, LLMs directly perform recommendation tasks, such as Tok-K recommendation and explanation generation), \colorbox{green!20}{$\langle$ \emph{Bridge LLMs and  RecSys} $\rangle$} (\textit{e.g.}, LLMs provide data augmentation for training recommendation models), and \colorbox{red!20}{$\langle$ \emph{LLM-based autonomous agent} $\rangle$} (\textit{e.g.}, LLMs simulate human-level user behaviors in RecSys \& Manage complex recommendations into sub-tasks).
}
\label{tab:prompting LLM4Rec}
\renewcommand\arraystretch{1.25}
\scalebox{0.8}
{
\begin{tabular}{ccccc}
\toprule
\multicolumn{1}{c|}{Paradigms} & \multicolumn{1}{c|}{Methods} & \multicolumn{1}{c|}{LLM Tasks} & \multicolumn{1}{c|}{LLM Backbones} & References \\ \bottomrule

\multicolumn{1}{c|}{\multirow{20}{*}{Prompting}} & \multicolumn{1}{c|}{\multirow{2}{*}{Conventional Prompting}} & \multicolumn{1}{c|}{\cellcolor{cyan!20}Text Summarization} & \multicolumn{1}{c|}{\cellcolor{cyan!20}ChatGPT} &
\cellcolor{cyan!20}\begin{tabular}[c]{@{}c@{}}
~\cite{liu2023chatgpt}
\end{tabular} \\ \hhline{~|~|-|-|-}

\multicolumn{1}{c|}{} & \multicolumn{1}{c|}{} & \multicolumn{1}{c|}{\multirow{1}{*}{\cellcolor{cyan!20}Relationship Extraction}} & \multicolumn{1}{c|}{\cellcolor{cyan!20}ChatGPT} &
\cellcolor{cyan!20}\begin{tabular}[c]{@{}c@{}}
~\cite{chen2023knowledge}
\end{tabular} \\ \hhline{~|-|-|-|-}

\multicolumn{1}{c|}{} & \multicolumn{1}{c|}{} & \multicolumn{1}{c|}{\cellcolor{cyan!20}} & \multicolumn{1}{c|}{\cellcolor{cyan!20}GPT-4} &
\cellcolor{cyan!20}\begin{tabular}[c]{@{}c@{}}
~\cite{he2023large}$^1$
\end{tabular} \\ \cline{4-5}
\multicolumn{1}{c|}{} & \multicolumn{1}{c|}{\multirow{13}{*}{In-context Learning (ICL)}} & \multicolumn{1}{c|}{\cellcolor{cyan!20}} & \multicolumn{1}{c|}{\cellcolor{cyan!20}ChatGPT} &
\cellcolor{cyan!20}\begin{tabular}[c]{@{}c@{}}
~\cite{liu2023chatgpt},~\cite{zhiyuli2023bookgpt},~\cite{he2023large}$^1$,~\cite{dai2023uncovering}$^2$,~\cite{hou2023large}$^3$,~\cite{wang2023rethinking}$^4$
\end{tabular} \\ \cline{4-5}
\multicolumn{1}{c|}{} & \multicolumn{1}{c|}{} & \multicolumn{1}{c|}{\cellcolor{cyan!20}} & \multicolumn{1}{c|}{\cellcolor{cyan!20}T5} &
\cellcolor{cyan!20}\begin{tabular}[c]{@{}c@{}}
~\cite{leszczynski2023generating},~\cite{wu2023towards}$^5$
\end{tabular} \\ \cline{4-5}
\multicolumn{1}{c|}{} & \multicolumn{1}{c|}{} & \multicolumn{1}{c|}{\cellcolor{cyan!20}\multirow{-4}{*}{
\begin{tabular}[c]{@{}c@{}}
Recommendation Tasks \\
(e.g., \textit{rating prediction}, \textit{top-K recommendation}, \\
\textit{conversational recommendation}, 
\textit{explanation generation}, etc.)
\end{tabular}
}} & \multicolumn{1}{c|}{\cellcolor{cyan!20}PaLM} &
\cellcolor{cyan!20}\begin{tabular}[c]{@{}c@{}}
~\cite{christakopoulou2023large},~\cite{sanner2023large}
\end{tabular} \\ \hhline{~|~|-|-|-}

\multicolumn{1}{c|}{} & \multicolumn{1}{c|}{} & \multicolumn{1}{c|}{\cellcolor{green!20}} & \multicolumn{1}{c|}{\cellcolor{green!20}GPT-4} &
\cellcolor{green!20}\begin{tabular}[c]{@{}c@{}}
~\cite{wang2023enhancing}
\end{tabular} \\ \cline{4-5}
\multicolumn{1}{c|}{} & \multicolumn{1}{c|}{} & \multicolumn{1}{c|}{\cellcolor{green!20}} & \multicolumn{1}{c|}{\cellcolor{green!20}ChatGPT} &
\cellcolor{green!20}\begin{tabular}[c]{@{}c@{}}
~\cite{wang2023enhancing},~\cite{liu2023first}$^6$,~\cite{wei2023llmrec}$^7$
\end{tabular} \\ \cline{4-5}
\multicolumn{1}{c|}{} & \multicolumn{1}{c|}{} & \multicolumn{1}{c|}{\cellcolor{green!20}\multirow{-3}{*}{Data Augmentation of RecSys}} & \multicolumn{1}{c|}{\cellcolor{green!20}GPT-3} &
\cellcolor{green!20}\begin{tabular}[c]{@{}c@{}}
~\cite{mysore2023large}
\end{tabular} \\ \hhline{~|~|-|-|-}

\multicolumn{1}{c|}{} & \multicolumn{1}{c|}{} & \multicolumn{1}{c|}{\cellcolor{green!20}} & \multicolumn{1}{c|}{\cellcolor{green!20}ChatGPT} &
\cellcolor{green!20}\begin{tabular}[c]{@{}c@{}}
~\cite{gao2023chat},~\cite{du2023enhancing}
\end{tabular} \\ \cline{4-5}
\multicolumn{1}{c|}{} & \multicolumn{1}{c|}{} & \multicolumn{1}{c|}{\cellcolor{green!20}} & \multicolumn{1}{c|}{\cellcolor{green!20}GPT-3} &
\cellcolor{green!20}\begin{tabular}[c]{@{}c@{}}
~\cite{lyu2023llmrec}
\end{tabular} \\ \cline{4-5}
\multicolumn{1}{c|}{} & \multicolumn{1}{c|}{} & \multicolumn{1}{c|}{\cellcolor{green!20}} & \multicolumn{1}{c|}{\cellcolor{green!20}GPT-2} &
\cellcolor{green!20}\begin{tabular}[c]{@{}c@{}}
~\cite{li2023gpt4rec}
\end{tabular} \\ \cline{4-5}
\multicolumn{1}{c|}{} & \multicolumn{1}{c|}{} & \multicolumn{1}{c|}{\cellcolor{green!20}\multirow{-4}{*}{Data Refinement of RecSys}} & \multicolumn{1}{c|}{\cellcolor{green!20}ChatGLM} &
\cellcolor{green!20}\begin{tabular}[c]{@{}c@{}}
~\cite{xi2023towards}$^8$
\end{tabular} \\ \hhline{~|~|-|-|-}

\multicolumn{1}{c|}{} & \multicolumn{1}{c|}{} & \multicolumn{1}{c|}{\cellcolor{green!20}API Call of RecSys \& Tools} & \multicolumn{1}{c|}{\cellcolor{green!20}ChatGPT} &
\cellcolor{green!20}\begin{tabular}[c]{@{}c@{}}
~\cite{wang2023recmind},~\cite{zhang2023graph}$^9$
\end{tabular} \\ \hhline{~|~|-|-|-}

\multicolumn{1}{c|}{} & \multicolumn{1}{c|}{} & \multicolumn{1}{c|}{\cellcolor{red!20}} & \multicolumn{1}{c|}{\cellcolor{red!20}GPT-4} &
\cellcolor{red!20}\begin{tabular}[c]{@{}c@{}}
~\cite{huang2023recommender}
\end{tabular} \\ \cline{4-5}
\multicolumn{1}{c|}{} & \multicolumn{1}{c|}{} & \multicolumn{1}{c|}{\cellcolor{red!20}\multirow{-2}{*}{User Behavior Simulation}} & \multicolumn{1}{c|}{\cellcolor{red!20}ChatGPT} &
\cellcolor{red!20}\begin{tabular}[c]{@{}c@{}}
~\cite{wang2023recagent}$^{10}$,~\cite{zhang2023generative}$^{11}$
\end{tabular} \\ \hhline{~|~|-|-|-}

\multicolumn{1}{c|}{} & \multicolumn{1}{c|}{} & \multicolumn{1}{c|}{\cellcolor{red!20}\multirow{1}{*}{Task Planning}} & \multicolumn{1}{c|}{\cellcolor{red!20}LLaMA} &
\cellcolor{red!20}\begin{tabular}[c]{@{}c@{}}
~\cite{feng2023large}
\end{tabular} \\ \hhline{~|-|-|-|-}

\multicolumn{1}{c|}{} & \multicolumn{1}{c|}{\multirow{3}{*}{Chain-of-thought (CoT)}} & \multicolumn{1}{c|}{\multirow{1}{*}{\cellcolor{cyan!20}Recommendation Tasks}} & \multicolumn{1}{c|}{\cellcolor{cyan!20}T5} &
\cellcolor{cyan!20}\begin{tabular}[c]{@{}c@{}}
~\cite{zhang2023recommendation}
\end{tabular} \\ \hhline{~|~|-|-|-}

\multicolumn{1}{c|}{} & \multicolumn{1}{c|}{} & \multicolumn{1}{c|}{\cellcolor{red!20}} & \multicolumn{1}{c|}{\cellcolor{red!20}GPT-4} &
\cellcolor{red!20}\begin{tabular}[c]{@{}c@{}}
~\cite{huang2023recommender}
\end{tabular} \\ \cline{4-5}
\multicolumn{1}{c|}{} & \multicolumn{1}{c|}{} & \multicolumn{1}{c|}{\cellcolor{red!20}\multirow{-2}{*}{Task Planning}} & \multicolumn{1}{c|}{\cellcolor{red!20}ChatGPT} &
\cellcolor{red!20}\begin{tabular}[c]{@{}c@{}}
~\cite{wang2023recmind}
\end{tabular} \\ \hhline{-|-|-|-|-}

\multicolumn{1}{c|}{\multirow{8}{*}{Prompt Tuning}} & \multicolumn{1}{c|}{\multirow{3}{*}{Hard Prompt Tuning}} & \multicolumn{1}{c|}{\cellcolor{cyan!20}Recommendation Tasks} & \multicolumn{1}{c|}{\cellcolor{cyan!20}GPT-2} & \multicolumn{1}{c}{
\cellcolor{cyan!20}\begin{tabular}[c]{@{}c@{}}
~\cite{li2023personalized}
\end{tabular}
} \\ \hhline{~|~|-|-|-}

\multicolumn{1}{c|}{} & \multicolumn{1}{c|}{} & \multicolumn{3}{c}{} \\
\multicolumn{1}{c|}{} & \multicolumn{1}{c|}{} & \multicolumn{3}{c}{\multirow{-2}{*}{
\begin{tabular}[c]{@{}c@{}}
ICL can be regarded as a subclass of prompt tuning, namely hard prompt tuning (see Section~\ref{subsec:hard_prompt_tuning} for explanations)
\end{tabular}
}}\\ \hhline{~|-|-|-|-}

\multicolumn{1}{c|}{} & \multicolumn{1}{c|}{\multirow{4}{*}{Soft Prompt Tuning}} & \multicolumn{1}{c|}{\cellcolor{cyan!20}} & \multicolumn{1}{c|}{\cellcolor{cyan!20}T5} &
\cellcolor{cyan!20}\begin{tabular}[c]{@{}c@{}}
~\cite{hua2023up5},~\cite{li2023prompt}
\end{tabular} \\ \cline{4-5}
\multicolumn{1}{c|}{} & \multicolumn{1}{c|}{} & \multicolumn{1}{c|}{\cellcolor{cyan!20}} & \multicolumn{1}{c|}{\cellcolor{cyan!20}GPT-2} &
\cellcolor{cyan!20}\begin{tabular}[c]{@{}c@{}}
~\cite{li2023personalized}
\end{tabular} \\ \cline{4-5}
\multicolumn{1}{c|}{} & \multicolumn{1}{c|}{} & \multicolumn{1}{c|}{\cellcolor{cyan!20}} & \multicolumn{1}{c|}{\cellcolor{cyan!20}PaLM} &
\cellcolor{cyan!20}\begin{tabular}[c]{@{}c@{}}
~\cite{christakopoulou2023large}
\end{tabular} \\ \cline{4-5}
\multicolumn{1}{c|}{} & \multicolumn{1}{c|}{} & \multicolumn{1}{c|}{\cellcolor{cyan!20}\multirow{-4}{*}{Recommendation Tasks}} & \multicolumn{1}{c|}{\cellcolor{cyan!20}M6} &
\cellcolor{cyan!20}\begin{tabular}[c]{@{}c@{}}
~\cite{cui2022m6}
\end{tabular} \\ \hhline{-|-|-|-|-}

\multicolumn{1}{c|}{\multirow{5}{*}{Instruction Tuning}} & \multicolumn{1}{c|}{\multirow{2}{*}{
\begin{tabular}[c]{@{}c@{}}
Full-model Tuning\\
with Prompt
\end{tabular}
}} & \multicolumn{1}{c|}{\cellcolor{cyan!20}} & \multicolumn{1}{c|}{\cellcolor{cyan!20}T5} &
\cellcolor{cyan!20}\begin{tabular}[c]{@{}c@{}}
~\cite{zhang2023recommendation},~\cite{kang2023llms}
\end{tabular} \\ \cline{4-5}
\multicolumn{1}{c|}{} & \multicolumn{1}{c|}{} & \multicolumn{1}{c|}{\cellcolor{cyan!20}\multirow{-2}{*}{Recommendation Tasks}} & \multicolumn{1}{c|}{\cellcolor{cyan!20}LLaMA} &
\cellcolor{cyan!20}\begin{tabular}[c]{@{}c@{}}
~\cite{chen2023palr},~\cite{zheng2023generative}
\end{tabular} \\ \hhline{~|-|-|-|-}

\multicolumn{1}{c|}{} & \multicolumn{1}{c|}{
\begin{tabular}[c]{@{}c@{}}
Parameter-efficient Model\\
Tuning with Prompt
\end{tabular}
} & \multicolumn{1}{c|}{\cellcolor{cyan!20}Recommendation Tasks} & \multicolumn{1}{c|}{\cellcolor{cyan!20}LLaMA} &
\cellcolor{cyan!20}\begin{tabular}[c]{@{}c@{}}
~\cite{bao2023tallrec}$^{12}$,~\cite{wu2023exploring},~\cite{ji2023genrec}$^{13}$\\
\newline\\
\end{tabular} \\ \toprule

\multicolumn{5}{l}{
Code Availability: 
\footnotesize$^1$\url{https://github.com/AaronHeee/LLMs-as-Zero-Shot-Conversational-RecSys},
\footnotesize$^2$\url{https://github.com/rainym00d/LLM4RS},
} \\
\multicolumn{5}{l}{
\footnotesize$^3$\url{https://github.com/RUCAIBox/LLMRank},
\footnotesize$^4$\url{https://github.com/RUCAIBox/iEvaLM-CRS},
\footnotesize$^5$\url{https://github.com/JacksonWuxs/PromptRec},
} \\ 
\multicolumn{5}{l}{  
\footnotesize$^6$\url{https://github.com/Jyonn/GENRE-requests},
\footnotesize$^7$\url{https://github.com/HKUDS/LLMRec},
} \\
\multicolumn{5}{l}{
\footnotesize$^8$\url{https://github.com/YunjiaXi/Open-WorldKnowledge-Augmented-Recommendation},
\footnotesize$^9$\url{https://github.com/jwzhanggy/Graph\_Toolformer},
} \\
\multicolumn{5}{l}{
\footnotesize$^{10}$\url{https://github.com/RUC-GSAI/YuLan-Rec},
\footnotesize$^{11}$\url{https://github.com/LehengTHU/Agent4Rec},
} \\
\multicolumn{5}{l}{
\footnotesize$^{12}$\url{https://anonymous.4open.science/r/LLM4Rec-Recsys},
\footnotesize$^{13}$\url{https://github.com/rutgerswiselab/GenRec}.
} \\

\toprule

\multicolumn{5}{l}{
\begin{tabular}[l]{@{}l@{}}
\footnotesize Note: some references with pre-trained LM backbones (\textit{e.g.}, GPT-2) are included since the corresponding methods are compared with LLM-based baselines.
\end{tabular}
}

\end{tabular}
}

\end{table*}

Apart from pre-training and fine-tuning paradigms, prompting serves as the latest paradigm for adapting LLMs to specific downstream tasks with the help of task-specific prompts.
A prompt refers to a text template that can be applied to the input of LLMs. For example, a prompt \emph{``The relation between \_ and \_ is \_.''} can be designed to deploy LLMs for relation extraction tasks. Prompting enables LLMs to unify different downstream tasks into language generation tasks, which are aligned to their objectives during pre-training~\cite{gao2020making}. 
Compared to pre-training and fine-tuning LLMs that require large task-specific datasets and costly parameter updates, prompting makes it possible to adapt LLMs to recommendation tasks in more lightweight manners, such as providing appropriate task instructions in natural language.
For instance, the popular ChatGPT retrieval plugin\footnote{https://github.com/openai/chatgpt-retrieval-plugin} serves as an API schema of prompting, which retrieves customized documents as prompts to the input of ChatGPT.
As highlighted in Table~\ref{tab:prompting LLM4Rec}, we categorize the insights of prompting LLMs for recommendations into three representative approaches, namely \emph{LLMs act as recommender}, \emph{Bridge LLMs and RecSys}, and \emph{LLM-based autonomous agent}, along with each subclass of prompting methods.

Recent research has actively explored prompting to facilitate the performance of LLMs for recommendations, advanced techniques like In-context Learning (ICL) and Chain-of-thought (CoT) are increasingly investigated to manually design prompts for various recommendation tasks.
In addition, prompt tuning serves as an additive technique of prompting, by adding prompt tokens to LLMs and then updating them based on task-specific recommendation datasets. 
More recently, instruction tuning that combines the pre-training $\&$ fine-tuning paradigm with prompting~\cite{wei2021finetuned} is explored to fine-tune LLMs over multiple recommendation tasks with instruction-based prompts, which enhances the \emph{zero-shot} performance of LLMs on unseen recommendation tasks.
Figure~\ref{fig:prompting} compares the representative methods corresponding to each of the aforementioned three prompting techniques of LLMs, in terms of the workflow of LLMs in recommender systems, input formation, and parameter update of LLMs (\textit{i.e.}, either tunable or frozen).
In this section, we will discuss the prompting, prompt tuning, and instruction tuning techniques in detail, for improving the performance of LLMs on recommendation tasks.
In summary, Table~\ref{tab:prompting LLM4Rec} categorizes the existing works according to the aforementioned three techniques, including the specific recommendation tasks and the LLM backbones considered in these works.

\subsection{Prompting}
\label{sec:prompting}
The key idea of prompting is to keep LLMs frozen (\textit{i.e.}, no parameters updates), and adapt LLMs to downstream tasks via task-specific prompts.
To recap the development of prompting strategies for adapting LLMs to downstream tasks,  early-stage conventional prompting methods mainly target unifying downstream tasks to language generation manners, such as text summarization, relation extraction, and sentiment analysis. 
Later on, ICL~\cite{brown2020language} emerges as a powerful prompting strategy that allows LLMs to learn new tasks (\textit{i.e.}, tasks with knowledge demanding objectives) based on contextual information. In addition, another up-to-date prompting strategy named CoT~\cite{wei2022chain} serves as a particularly effective method for prompting LLMs to address downstream tasks with complex reasoning.

\subsubsection{Conventional Prompting}
There are two major approaches for prompting pre-trained language models to improve the performance on specific downstream tasks. 
One approach is prompt engineering, which generates prompt by emulating text that language models encountered during pre-training (\textit{e.g.}, text in NLP tasks). 
This allows pre-trained language models to unify downstream tasks with unseen objectives into language generation tasks with known objectives.
For instance, Liu \textit{et al.}~\cite{liu2023chatgpt} consider prompting ChatGPT to format the review summary task in recommendations into generic language generation task of text summarization, using a prompt \emph{``Write a short sentence to summarize''}.
Another approach is few-shot prompting, where a few input-output examples (\textit{i.e.}, shots) are provided to prompt and guide pre-trained language models to generate desired output for specific downstream tasks.

Due to the huge gap between language generation tasks (\textit{i.e.}, the pre-training objectives of LLMs) and downstream recommendation tasks, most conventional prompting methods have only shown limited applications in specific recommendation tasks that have similar nature to language generation tasks, such as the review summary of users~\cite{liu2023chatgpt} and the relation labeling between items~\cite{chen2023knowledge}.

\subsubsection{In-context Learning (ICL)}
\begin{figure}[!t]
    \centering
    \includegraphics[width=0.85\linewidth]{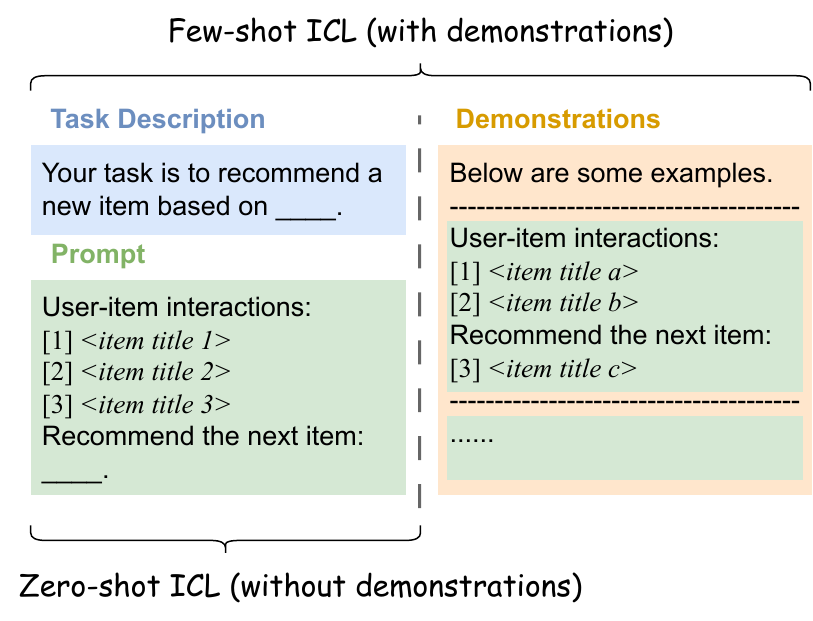}
    \caption{A brief template of zero-shot ICL and few-shot ICL for recommendation tasks.}
    \label{fig:ICL_template}    
\end{figure}

Alongside the introduction of GPT-3~\cite{brown2020language}, ICL is proposed as an advanced prompting strategy, which significantly boosts the performance of LLMs on adapting to many downstream tasks. Gao \textit{et al.}~\cite{gao2020making} attribute the success of ICL in prompting LLMs for downstream tasks to two designs: prompt and in-context demonstrations.
In other words, the key innovation of ICL is to elicit the in-context ability of LLMs for learning (new or unseen) downstream tasks from context during the inference stage.
In particular, two settings proposed in ICL are prevalently leveraged for prompting LLMs for RecSys. 
One is the few-shot setting, in which a few demonstrations with contexts and desired completions of the specific downstream tasks are provided along with prompts. 
The other is the zero-shot setting, where no demonstrations will be given to LLMs but only natural language descriptions of the specific downstream tasks are appended to the prompt.
As shown in Figure~\ref{fig:ICL_template}, a brief template of zero-shot ICL and few-shot ICL for recommendation tasks is provided.

Many existing works consider both few-shot ICL and zero-shot ICL settings at the same time to compare their performance under the same recommendation tasks.
Typically, few-shot ICL can outperform zero-shot ICL since additional in-context demonstrations are provided to LLMs.
Despite the reduction in performance, zero-shot ICL entirely relieves the requirement of task-specific recommendation datasets to form in-context demonstrations and can be suitable for certain tasks like conversational recommendations, where users are not likely to provide any demonstration to LLMs. For example, Wang \textit{et al.}~\cite{wang2023rethinking} prompt ChatGPT for conversational recommendations with a zero-shot ICL template containing two parts: a text description of conversational recommendation tasks (\textit{e.g.}, \emph{``Recommend items based on user queries in the dialogue.''}), and a format guideline in natural languages, such as \emph{``The output format should be $\langle$no.$\rangle$ $\langle$item title$\rangle$.''}, making the recommendation results easier to parse.

To adapt LLMs to recommendation tasks via ICL, a straightforward approach is to teach LLMs to act as recommenders.
For instance, Liu \textit{et al.}~\cite{liu2023chatgpt} employ ChatGPT and propose separate task descriptions tailored to different recommendation tasks, including top-K recommendation, rating prediction, and explanation generation, to perform ICL based on corresponding input-output examples of each recommendation task. For instance, the user rating history is given as an example for rating prediction tasks.
Similarly, other existing works propose their distinct insights into designing the in-context demonstrations for better recommendation performance.
For example, a text description of role injection, such as \emph{``You are a book rating expert.''}, is proposed in~\cite{zhiyuli2023bookgpt} to augment the in-context demonstrations, which prevents LLMs from refusing to complete the recommendation tasks (\textit{e.g.}, LLMs sometimes respond with \emph{``As a language model, I don't have the ability to recommend ...''} for recommendation tasks).

Apart from teaching LLMs to directly act as RecSys, ICL is also leveraged to bridge LLMs and conventional recommendation models. For example, a framework named Chat-Rec~\cite{gao2023chat} is proposed to bridge ChatGPT and traditional RecSys via ICL, where ChatGPT learns to receive candidate items from traditional RecSys and then refines the final recommendation results.
What’s more, Zhang~\cite{zhang2023graph} designs a textual API call template for external graph reasoning tools and successfully teaches ChatGPT to use those templates through ICL to access the graph-based recommendation results generated by the external tools.
More recently, LLM-based autonomous agents have been explored to simulate user behaviors in RecSys, such as InteRecAgent~\cite{huang2023recommender}, RecAgent~\cite{wang2023recagent}, and Agent4Rec~\cite{zhang2023generative}, by equipping LLMs with memory and action modules. In particular, few-show ICL methods are designed to connect LLMs with these external modules, enabling LLMs to interact with RecSys and simulate user behaviors like chatting and posting.

\subsubsection{Chain-of-thought (CoT) Prompting}
Although ICL has shown great effectiveness in prompting LLMs for downstream tasks with in-context demonstrations, recent studies indicate that LLMs still have limited performance in reasoning-heavy tasks~\cite{wei2022chain}. 
More specifically, by prompting LLMs with in-context examples of input-output pairs, the answers directly generated by LLMs often suffer from 
missing one or a few intermediate reasoning steps 
in multi-step problems like mathematical equations, leading to a broken reasoning logic that causes errors in the subsequent reasoning steps (\textit{i.e.}, ``one-step missing errors''~\cite{wei2022chain}). Similar multi-step problems also exist in RecSys, such as the multi-step reasoning of user preferences based on the multi-turn dialogues in conversational recommendations. 
To address such limitations, CoT offers a special prompting strategy to enhance the reasoning ability of LLMs, by annotating intermediate reasoning steps to prompt. 
This enables LLMs to break down complicated decision-making processes and generate the final output with step-by-step reasoning.

Considering the suitable prompting strategies for adapting LLMs to various downstream tasks with complex reasoning, Zhao \textit{et al.}~\cite{zhao2023survey} discuss the combination of ICL and CoT prompting under two major settings: zero-shot CoT and few-shot CoT.
By inserting tricky texts such as ``\emph{Let’s think step by step}" and ``\emph{Therefore, the answer is}" to prompt, zero-shot CoT leads LLMs to generate task-specific reasoning steps independently, without providing any task-relevant instruction or grounding example.
As for few-shot CoT, task-specific reasoning steps are manually designed for each demonstration in ICL, where the original input-output examples are augmented to input-CoT-output manners. Besides, CoT can also augment the task descriptions in ICL demonstrations, by adding interpretable descriptions of reasoning steps based on task-specific knowledge.

In practice, the design of appropriate CoT reasoning steps highly depends on the contexts and objectives of the specific recommendation tasks. 
For example, a simple CoT template ``\emph{Please infer the preference of the user and recommend suitable items.}" is proposed to guide LLMs to first infer the user's explicit preference and then generate final recommendations~\cite{zhang2023recommendation}.
Below, we present a preliminary idea of CoT prompting, through an example in the context of e-commerce recommendations.

\begin{center}
    \begin{tabularx}{0.45\textwidth}{X}
        \emph{\textbf{[CoT Prompting]} Based on the user purchase history, \textbf{let's think step-by-step}. \textbf{First}, please infer the user's high-level shopping intent. \textbf{Second}, what items are usually bought together with the purchased items? \textbf{Finally}, please select the most relevant items based on the shopping intent and recommend them to the user.}
    \end{tabularx}
\end{center}

\noindent More recently, studies like InteRecAgent~\cite{huang2023recommender} and RecMind~\cite{wang2023recmind} have employed CoT prompting, enabling LLMs to act as agents and manage complex recommendations into sub-tasks by generating plans for utilizing external tools.

Beyond the RecSys field, a recent research~\cite{yao2023beyond} has revealed the great effectiveness of adopting CoT prompting to facilitate the graph reasoning ability of LLMs (T5 particularly) by modeling the reasoning steps as nodes and connecting the reasoning paths as edges instead of a sequential chain. 
We believe that similar ideas can be potentially transferred and contribute to the CoT prompting for RecSys, based on the fact that recommendation tasks can be considered as a special case of link prediction problems in graph learning.

\subsection{Prompt Tuning}
\label{sec:prompt tuning}
In contrast to manually prompting LLMs for downstream tasks (\textit{e.g.}, manually generate task-specific prompt in natural language), prompt tuning serves as an additive technique of prompting, which adds new prompt tokens to LLMs and optimizes the prompt based on the task-specific dataset.
Generally, prompt tuning requires less task-specific knowledge and human effort than manually designing prompts for specific tasks and only involves minimal parameter updates of the tunable prompt and the input layer of LLMs. 
For example, AutoPrompt~\cite{shin2020autoprompt} takes the step of decomposing prompt into a set of vocabulary tokens, and finding the suitable tokens to language models via gradient-based search with respect to the performance on specific tasks.

According to the definition, prompts can be either discrete (\textit{i.e.}, hard) or continuous (\textit{i.e.}, soft) that guide LLMs to generate the expected output~\cite{dong2022survey}. Thus, we categorize prompt tuning strategies for prompting LLMs for RecSys into hard prompt tuning and soft prompt tuning, as illustrated below.

\subsubsection{Hard Prompt Tuning} \label{subsec:hard_prompt_tuning}
Hard prompt tuning is to generate and update discrete text templates of prompt (\textit{e.g.}, in natural language), for prompting LLMs to specific downstream tasks. 
Dong et al.~\cite{dong2022survey} argue that ICL can be considered as a subclass of hard prompt tuning and regard the in-context demonstrations in ICL as a part of the prompt. From this perspective, ICL performs hard prompt tuning for prompting LLMs to downstream recommendation tasks by refining prompts in natural language based on task-specific recommendation datasets.
Despite the effectiveness and convenience of generating or refining natural language prompts for downstream recommendation tasks, hard prompt tuning inevitably faces the challenge of discrete optimization, which requires laborious trial and error to discover the vast vocabulary space in order to find suitable prompts for specific recommendation tasks.

\subsubsection{Soft Prompt Tuning}
In contrast to discrete prompt, soft prompt tuning employs continuous vectors as prompt (\textit{e.g.}, text embeddings), and optimizes the prompt based on task-specific datasets, such as using gradient methods to update the prompt with respect to a recommendation loss. In LLMs, soft prompt tokens are often concatenated to the original input tokens at the input layer (\textit{e.g.}, tokenizer).
During soft prompt tuning, only the soft prompt and minimal parameters at the input layer of LLMs will be updated. 

To improve the recommendation performance of LLMs, some existing works combine advanced feature extraction and representation learning methods to better capture and embed task-specific information in RecSys into soft prompts.
For instance, Wu \textit{et al.}~\cite{wu2022personalized} apply contrastive learning to capture user representations and encode them into prompt tokens, and Wang \textit{et al.}~\cite{wang2022towards} and Guo \textit{et al.}~\cite{guo2023automated} share the similar idea of encoding mutual information in cross-domain recommendations into soft prompt.
In addition to directly embedding task-specific information into soft prompts, soft prompts can also be learned based on task-specific datasets.
For example, randomly initialized soft prompts are adopted to guide T5 to generate desired recommendation results~\cite{hua2023up5}, where the soft prompt is optimized in an end-to-end manner with respect to a recommendation loss based on the T5 output.
Compared to the hard prompt, the soft prompt is more feasible for tuning on continuous space but at a cost of explainability~\cite{hua2023up5}. 
In other words, compared to task-specific hard prompt in a natural language like \emph{``Your task is to recommend ...''}, the relationships between the specific downstream tasks and the soft prompt written in continuous vectors are not interpretable to humans.

\subsection{Instruction Tuning} 
\label{sec:instruction tuning}
Although prompting LLMs has demonstrated remarkable few-shot performance on unseen downstream tasks, recent studies demonstrated that prompting strategies have much poorer zero-shot ability~\cite{wei2021finetuned}.
To address the limitations, instruction tuning is proposed to fine-tune LLMs over multiple task-specific prompts.
In other words, instruction tuning possesses features of both prompting and pre-training $\&$ fine-tuning paradigms. 
This helps LLMs gain better capabilities of exactly following prompts as instructions for diverse downstream tasks, which hence contributes to the enhanced zero-shot performance of LLMs on unseen tasks by accurately following new task instructions. 
The key insight of instruction tuning is to train LLMs to follow prompts as task instructions, rather than to solve specific downstream tasks.
More specifically, instruction tuning can be divided into two stages: ``instruction'' (\textit{i.e.}, prompt) generation and model ``tuning'', since the straightforward idea of instruction tuning is the combination of prompting and fine-tuning LLMs.
\begin{itemize}
    \item \textbf{\emph{Instruction (Prompt) Generation  Stage.}} 
    Formally, instruction tuning introduces a format of instruction-based prompt in natural language, which consists of task-oriented input (\textit{i.e.}, task descriptions based on task-specific dataset) and desired target (\textit{i.e.}, corresponding output based on task-specific dataset) pairs. Considering the instruction tuning of LLMs for downstream recommendation tasks, Zhang \textit{et al.}~\cite{zhang2023recommendation} propose a recommendation-oriented instruction template, including user preferences, intentions, and task forms, which serves as a common template for generating instructions for various recommendation tasks. More directly, three-part instruction templates in the form of "task description-input-output" are used in~\cite{bao2023tallrec,chen2023palr} to generate instructions based on task-specific recommendation datasets.

    \item \textbf{\emph{Model Tuning Stage.}} The second stage is to fine-tune LLMs over multiple aforementioned instructions for downstream tasks,  where we categorize the existing works on RecSys, as shown in Table~\ref{tab:prompting LLM4Rec}, according to the LLMs fine-tuning manners: full-model tuning and parameter-efficient model tuning (see Section~\ref{sec:fine-tuning} for explanations), since basically the same principles of fine-tuning LLMs are adopted in this stage. For example, Bao \textit{et al.}~\cite{bao2023tallrec} utilize LoRA to make the instruction tuning of LLaMA more lightweight for downstream recommendation tasks.
\end{itemize}

In addition to textual data in RecSys, instruction tuning has recently been explored to enhance the graph understanding ability of LLMs for recommendation tasks. 
In particular, Wu \textit{et al.}~\cite{wu2023exploring} propose an LLM-based prompt constructor to encode the paths of nodes (\textit{e.g.}, candidate items) and edges (\textit{e.g.}, relationships between items) in behavior graphs into natural language descriptions, which is subsequently used for instruction tuning an LLM-based recommender based on task-specific datasets.

\section{Future Directions}
\label{sec:future_work}
In this survey, we have comprehensively reviewed the recent advanced techniques for LLM-enhanced recommender systems.
Since the adaption of LLMs to recommender systems is still in an early stage, there are still many challenges, which are also the opportunities. 
In this section, we discuss some potential future directions in this field.

\subsection{Hallucination Mitigation}  
Although LLMs are used in various fields, a significant challenge is the phenomenon of '\emph{hallucination}', where language models generate outputs that are plausible-sounding but factually incorrect or not referable in the input data~\cite{manakul2023selfcheckgpt,mckenna2023sources}. 
For instance, considering a scenario where you are seeking today's news events, the LLMs erroneously recommend/generate news that, in fact, does not exist.
The causes of this problem are manifold such as source-reference divergence existing in the dataset, and training\&modeling choices of neural network models~\cite{ji2023survey}.
Moreover, the hallucination issue poses severe threats to users and society, especially in high-stakes recommendation scenarios such as medical recommendations or legal advice, where the dissemination of incorrect information can have severe real consequences. 
To address such issues, employing factual knowledge graphs as supplementary factual knowledge during the training and inference stages of LLMs for RecSys is promising to mitigate the hallucination problem. 
In addition, the model's output stage can be scrutinized to verify the accuracy and factuality of the produced content.

\subsection{Trustworthy Large Language Models for Recommender Systems}
The development of LLMs for RecSys has brought significant benefits to humans, including economic value creation, time and effort savings, and social benefits.
However, these data-driven LLMs for RecSys might also pose serious threats to users and society~\cite{zhuo2023exploring,fan2022comprehensive,liu2022trustworthy}, due to unreliable decision-making,  unequal treatment of various consumers or producers, a lack of transparency and explainability, and privacy issues stemming from the extensive use of personal data for customization, among other concerns. 
As a result, there is an increasing concern about the issue of trustworthiness in LLMs for RecSys to mitigate the negative impacts and enhance public trust in LLM-based RecSys techniques. 
Thus, it is desired to achieve trustworthiness in LLMs for RecSys from four of the most crucial dimensions, including \emph{Safety\&Robustness, Non-discrimination\&Fairness, Explainability}, and \emph{Privacy}.

\subsubsection{Safety\&Robustness} 
LLMs have been proven to advance recommender systems in various aspects, but they are also highly vulnerable to adversarial perturbations (\textit{i.e.}, minor changes in the input) that can compromise the safety and robustness of their uses in safety-critical applications~\cite{zhang2023certified,zhuo2023exploring}. 
These vulnerabilities towards noisy inputs are frequently carried out with malicious intent, such as to gain unlawful profits and manipulate markets for specific products~\cite{fan2021attacking,chen2022knowledge,fan2023adversarial,fan2023jointly}. 
Therefore, it is crucial to ensure that the output of LLMs for recommender systems is stable given small changes in the LLMs' input. 
In order to enhance model safety and robustness, GPT-4 integrates safety-related prompts during reinforcement learning from human feedback (RLHF)~\cite{GPT4report}. 
However, the RLHF method requires a significant number of experts for manual labeling, which might not be feasible in practice. An alternative solution might involve the automatic pre-processing of prompts designed for recommender tasks before input to LLMs. This could include pre-processing for malicious prompts or standardizing prompts with similar purposes to have the same final input, thus potentially improving safety and robustness. 
In addition, as one of the representative techniques,  adversarial training~\cite{tang2019adversarial} can be used to improve the robustness of LLM-based recommender systems.

\subsubsection{Non-discrimination\&Fairness}
LLMs, trained on vast datasets, often inadvertently learn and perpetuate biases and stereotypes in the human data that will later reveal themselves in the recommendation results. This phenomenon can lead to a range of adverse outcomes, from the propagation of stereotypes to the unfair treatment of certain user groups~\cite{chen2023fairly,zhang2022fairness,liu2020does}. 
For instance, in the context of recommender systems, these biases can manifest as discriminatory recommendations, where certain items are unfairly promoted or demoted based on these learned biases. 
More recently, a few studies such as FaiRLLM~\cite{zhang2023chatgpt} and UP5~\cite{hua2023up5} explore the fairness problem in recommender systems brought by LLMs, which only focus on user-side and item generation task.
Concurrently, Hou \textit{et al.}~\cite{hou2023large} guide LLMs with prompts to formalize the recommendation task as a conditional ranking task to improve item-side fairness. 
However, studies on non-discrimination and fairness in LLMs for RecSys are at a preliminary stage, further research is still needed.

\subsubsection{Explainability}
Owing to privacy and security considerations, certain companies and organizations choose not to open-source their advanced LLMs, such as ChatGPT and GPT-4,  indicating that the architectures and parameters of these LLMs for RecSys are not publicly available for the public to understand their complex internal working mechanisms.
Consequently, LLMs for RecSys can be treated as the 'black box', complicating the process for users trying to comprehend why a specific output or recommendation was produced.
Recently, Bills \textit{et al.}~\cite{bills2023language} try to use GPT-4 to generate natural language descriptions to explain the neuronal behavior in the GPT-2 model. 
While this study is foundational, it also introduces fresh perspectives for comprehending the workings of LLMs. Neurons exhibit intricate behaviors that may not be easily encapsulated through simple natural language. 
To this end, efforts should be made to understand how LLMs for RecSys function, so as to enhance the explainability of LLM-based recommender systems.

\subsubsection{Privacy}
Privacy is a paramount concern when it comes to LLMs for RecSys.
The reasons for this are multifold.  
On the one hand, the success of LLMs for recommender systems highly depends on large quantities of data that are collected from a variety of sources, such as social media and books. 
Users' sensitive information (\textit{e.g.}, email and gender) contained in data is likely to be used to train modern LLMs for enhancing prediction performance and providing personalized experiences, leading to the risk of leaking users' private information.
On the other hand, these systems often handle sensitive user data, including personal preferences, online behaviors, and other identifiable information. 
If not properly protected, this data could be exploited, leading to breaches of privacy. 
Therefore, ensuring the privacy and security of this data is crucial.
Carlini \textit{et al.}~\cite{carlini2021extracting} show that LLMs might reveal some uses' real identity or private information when generating text. 
Recently, Li \textit{et al.}~\cite{li2023privacy} introduce RAPT that allows users to customize LLMs with their private data based on prompt tuning. It provides a direction on how to protect user privacy at LLMs for RecSys. 

Notably, concurrent to the recent advancement of federated learning~\cite{tong2023federated} for facilitating data privacy in recommender systems~\cite{yang2020federated,zhang2022efficient}, LLMs have brought distinctive opportunities to the interplay between data privacy and federated learning~\cite{dong2023towards}. For instance, Zhuang et al.~\cite{zhuang2023foundation} systematically review the remarkable capabilities of LLMs as foundation models for federated learning, where LLMs are leveraged as controllers to seamlessly connect distributed devices. In particular, such scalable frameworks empowered by LLMs support the availability of federated learning on distributed data sources, which guarantees more privacy-preserving recommender systems by enabling localized learning and data privacy in a decentralized manner.

\subsection{Vertical Domain-Specific LLMs for Recommender Systems}
General LLMs, such as ChatGPT, whose powerful generation and inference capabilities make them a universal tool in various areas.
Vertical domain-specific LLMs are LLMs that have been trained and optimized for a specific domain or industry,
such as health~\cite{nastasi2023does} and finance~\cite{wu2023bloomberggpt}. Compared to general LLMs for RecSys, vertical domain-specific LLM-empowered RecSys are more focused on the knowledge and skills of a particular domain and have a higher degree of domain expertise and practicality.
Instead of sifting through irrelevant information, users can focus on content that is directly aligned with their work or personalized preferences. 
By providing tailored recommendations, vertical domain-specific LLMs for RecSys can save professionals a significant amount of time. 
More recently, existing works have presented vertical domain-specific LLMs that cover a wide range of areas, such as medical care~\cite{zhang2023huatuogpt,xiong2023doctorglm}, law~\cite{nguyen2023brief,huang2023lawyer}, and finance~\cite{yang2023fingpt}. 
Due to trained specifically, these vertical domain-specific LLMs can better understand and process domain-specific knowledge, terminology and context. 
Yet the requirement for vast amounts of domain-specific data to train these models poses significant challenges in data collection and annotation. 
As such, constructing high-quality domain datasets and using suitable tuning strategies for specific domains are necessary steps in the development of vertical domain-specific LLMs for RecSys.
In particular, Jin \textit{et al.}~\cite{jin2023amazon} propose a multilingual dataset named Amazon-M2 as a new setting of session-based recommendations from Amazon (\textit{i.e.}, sessions containing the interacted items of users) and inspire the opportunities to leverage LLMs as RecSys to learn on session graphs with multilingual and textual data, such as item (node) attributes including product titles, prices, and descriptions across session graphs of users from different locales (multilingual).

\subsection{Users\&Items Indexing}
Recent research suggests that LLMs may not perform well when dealing with long texts in RecSys, as it can be difficult to effectively capture user-item interaction information in long texts~\cite{hou2023large}.  
On the other hand,  user-item interactions (e.g., click, like, and subscription) with unique identities (i.e., discrete IDs) in recommender systems contain rich collaborative knowledge and make great contributions to understanding and predicting user preferences, encompassing both explicit actions like ratings and reviews, as well as implicit behaviors like browsing history or purchase data. 
Several studies, including InstructRec~\cite{zhang2023recommendation}, PALR~\cite{chen2023palr}, GPT4Rec~\cite{li2023gpt4rec} and UP5~\cite{hua2023up5}, have attempted to utilize user-item history interaction information as text prompts inputted into LLMs (\textit{e.g.}, ChatGPT) in order to make recommendations. To address the long text problem, one possible solution is to perform user and item indexing for learning collaborative knowledge by incorporating user-item interactions. 
Therefore, rather than merely using text formats to represent users and items,
advanced methods for indexing users\&items are desired to build LLM-based recommender systems.

\subsection{Fine-tuning Efficiency} 
In the application of LLMs to RecSys, fine-tuning refers to the process of adapting a pre-trained LLM to a specific task or domain, such as recommending movies~\cite{chen2023palr} or books~\cite{bao2023tallrec}. This process allows the model to leverage the general language understanding capabilities learned during pre-training while specializing its knowledge to the task at hand. However, fine-tuning can be computationally expensive, particularly for very large models and large datasets in recommender systems. Therefore, improving the efficiency of fine-tuning is a key challenge. 
In this case, Fu \textit{et al.}~\cite{fu2023exploring} use adapter modules, which are small, plug-in neural networks that can be optimized separately from the main model, to achieve parameter-efficient transfer learning. 
However, the current adapter tuning techniques for RecSys fall slightly behind full-model fine-tuning when it comes to cross-platform image recommendation. The exploration of adapter tuning effects for multi-modal (\textit{i.e.}, both text and image) RecSys is a potential future direction. 
In addition, given that most typical adapter tuning does not help to speed up the training process in practice, it is important to explore effective optimization techniques to reduce the computational cost and time for RecSys through end-to-end training.

\subsection{Data Augmentation}
Most conventional studies in the recommender systems domain rely on real data-driven research, founded on the collection of user behavior data via user interaction in digital platforms or through the recruitment of annotators. Nonetheless, these approaches appear to be resource-intensive and may not be sustainable in the long term.
The quality and variety of the input data directly influence the performance and versatility of the models. 
With the aim to overcome the shortcomings of real data-centric studies, Wang \textit{et al.}~\cite{wang2023recagent} introduce RecAgent, a simulation paradigm for recommender systems based on LLMs, which includes a user module for browsing and communication on the social media, and a recommender module for providing search or recommendation lists. 
Additionally, LLM-Rec~\cite{lyu2023llmrec} incorporates four prompting strategies to improve personalized content recommendations, which demonstrates through experiments that diverse prompts and input augmentation techniques can enhance recommendation performance. 
Therefore, rather than solely deploying LLMs as recommender systems, utilizing them for data augmentation to bolster recommendations emerges as a promising strategy in the future.

\section{Conclusion}
\label{Conclusion}
As one of the most advanced AI techniques, LLMs have achieved great success in various applications, such as molecule discovery and finance,  owing to their remarkable abilities in language understanding and generation, powerful generalization and reasoning skills, and prompt adaptation to new tasks and diverse domains. 
Similarly, increasing efforts have been made to revolutionize recommender systems with LLMs, so as to provide high-quality and personalized suggestion services. 
Given the rapid evolution of this research topic in recommender systems, there is a pressing need for a systematic overview that comprehensively summarizes the existing LLM-empowered recommender systems.  
To fill the gap, in this survey, we have provided a comprehensive overview of LLM-empowered RecSys from  \emph{pre-training\(\&\)fine-tuning} and \emph{prompting} paradigms, so as to provide researchers and practitioners in relevant fields with an in-depth understanding.
Nevertheless, the current research on LLMs for RecSys is still in its early stage which calls for more systematic and comprehensive studies of LLMs in this field. 
Therefore, we also discussed some potential future directions in this field.

\section*{Acknowledgments}

The research described in this paper has been partly supported by NSFC (project no. 62102335),  General Research Funds from the Hong Kong Research Grants Council (Project No.: PolyU 15200021, 15207322, and 15200023), internal research funds from The Hong Kong Polytechnic University (project no. P0036200, P0042693, P0048625, P0048752), Research Collaborative Project No. P0041282, and SHTM Interdisciplinary Large Grant (project no. P0043302). 
Xiangyu Zhao was supported by APRC-CityU New Research Initiatives (No.9610565, Start-up Grant for New Faculty of City University of Hong Kong), 
SIRG-CityU Strategic Interdisciplinary Research Grant (No.7020046, No.7020074), HKIDS Early Career Research Grant (No.9360163), and Ant Group (CCF-Ant Research Fund, Ant Group Research Fund).

\bibliographystyle{IEEEtran}

\bibliography{reference}

{\begin{IEEEbiography}[{\includegraphics[width=1in,height=1.25in,clip,keepaspectratio]{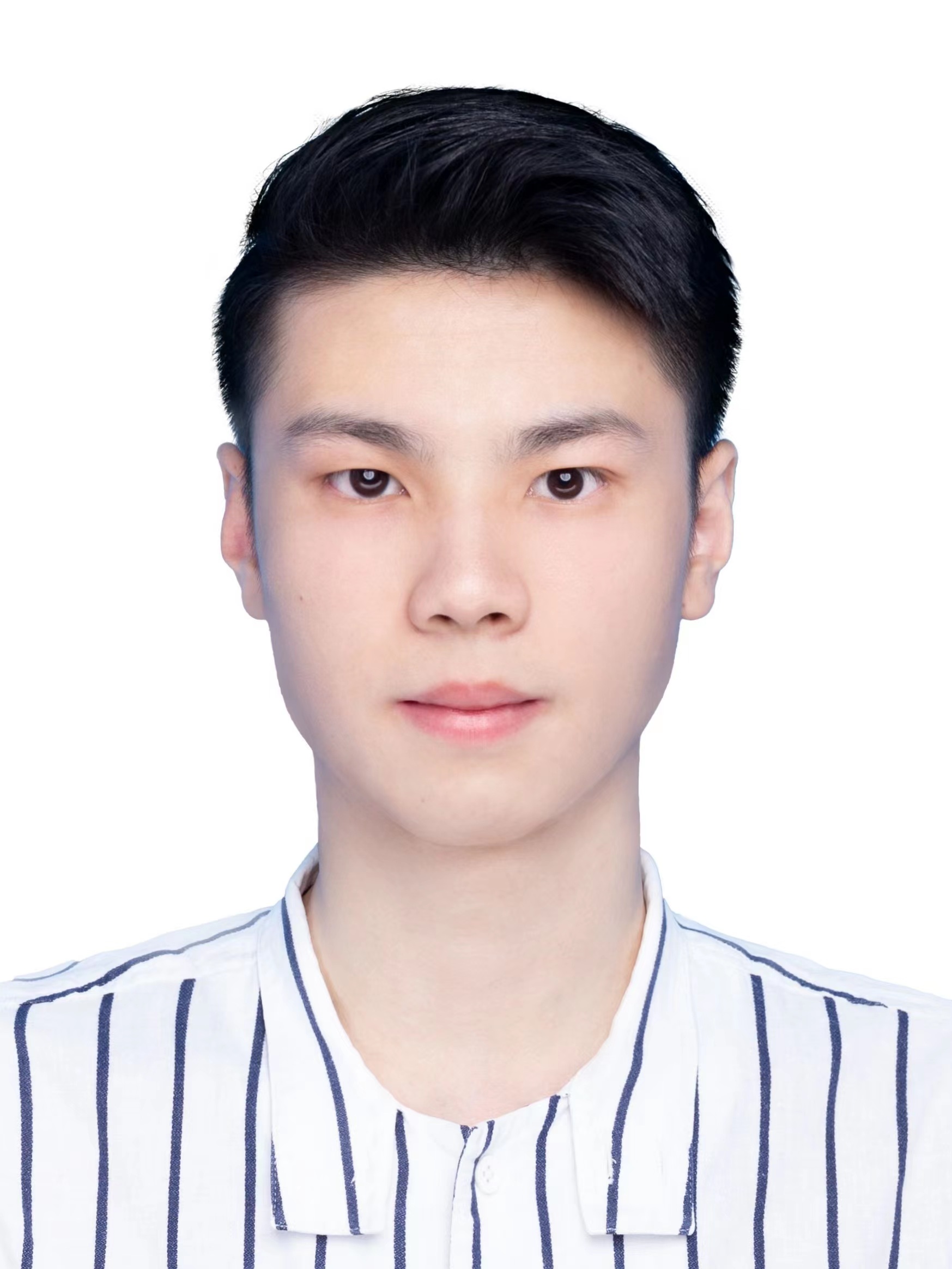}}]
{Zihuai Zhao} is currently a PhD student of the Department of Computing (COMP), Hong Kong Polytechnic University (PolyU), under the supervision of Prof. Qing Li and Dr. Wenqi Fan. Before joining the PolyU, he received both a Master’s degree (MPhil in Electrical Engineering) and a Bachelor’s degree (B.Eng. (Hons) in Electrical Engineering) from the University of Sydney in 2023 and 2020, respectively. His research interest covers Recommender Systems, Natural Language Processing, and Deep Reinforcement Learning. He has published innovative works in top-tier journals such as IoT-J. For more information, please visit https://scofizz.github.io/.

\end{IEEEbiography}

{\begin{IEEEbiography}[{\includegraphics[width=1in,height=1.25in,clip,keepaspectratio]{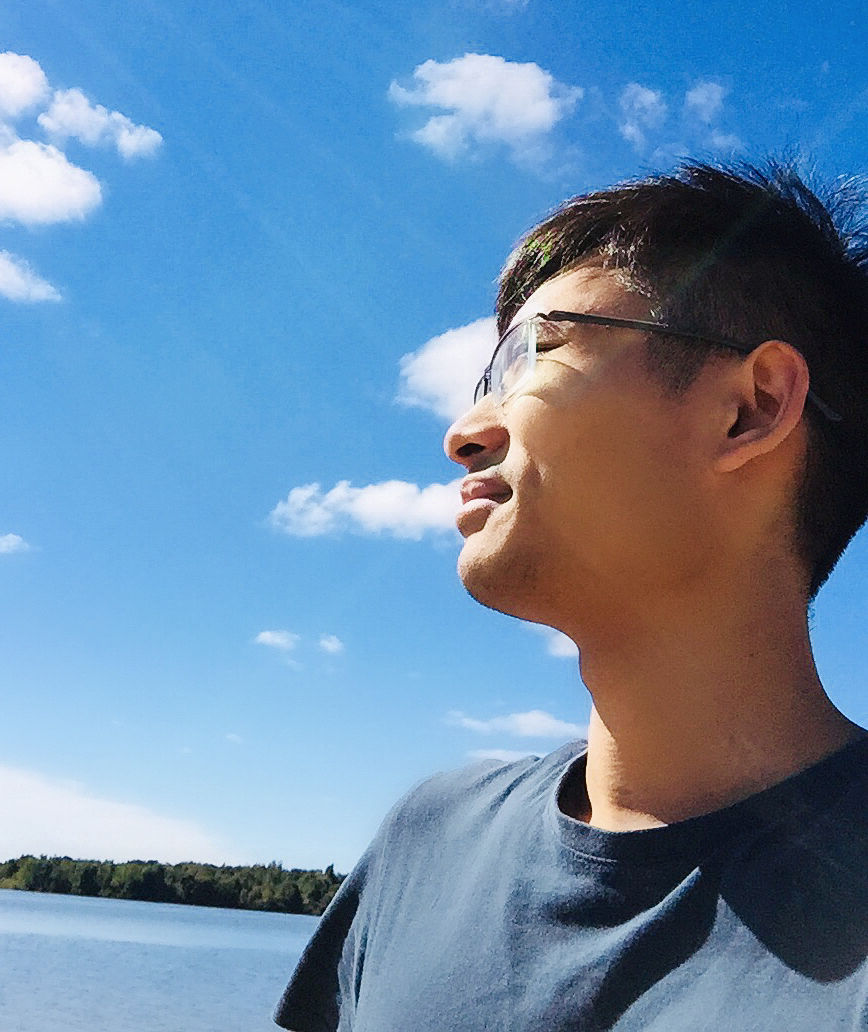}}]{Wenqi Fan} is an assistant professor of the Department of Computing (COMP) and the Department of Management and Marketing (MM) at The Hong Kong Polytechnic University (PolyU). He received his Ph.D. degree from the City University of Hong Kong (CityU) in 2020.
From 2018 to 2020, he was a visiting research scholar at Michigan State University (MSU). 
His research interests are in the broad areas of machine learning and data mining, with a particular focus on Recommender Systems, Graph Neural Networks, and Trustworthy Recommendations. He has published innovative papers in top-tier journals and conferences such as  TKDE, TIST, KDD, WWW, ICDE, NeurIPS, ICLR, SIGIR, IJCAI, AAAI, RecSys, WSDM, etc. 
He serves as top-tier conference (Area/Senior) Program Committee members and session chairs (e.g., ICML, ICLR, NeurIPS, KDD, WWW, AAAI, IJCAI, WSDM, EMNLP, ACL,  etc.), and journal reviewers (e.g., TKDE, TIST, TKDD, TOIS, TAI, etc.). 
More information about him can be found at https://wenqifan03.github.io.

\end{IEEEbiography}

\vspace{-24pt}
{\begin{IEEEbiography}[{\includegraphics[width=1in,height=1.25in,clip,keepaspectratio]{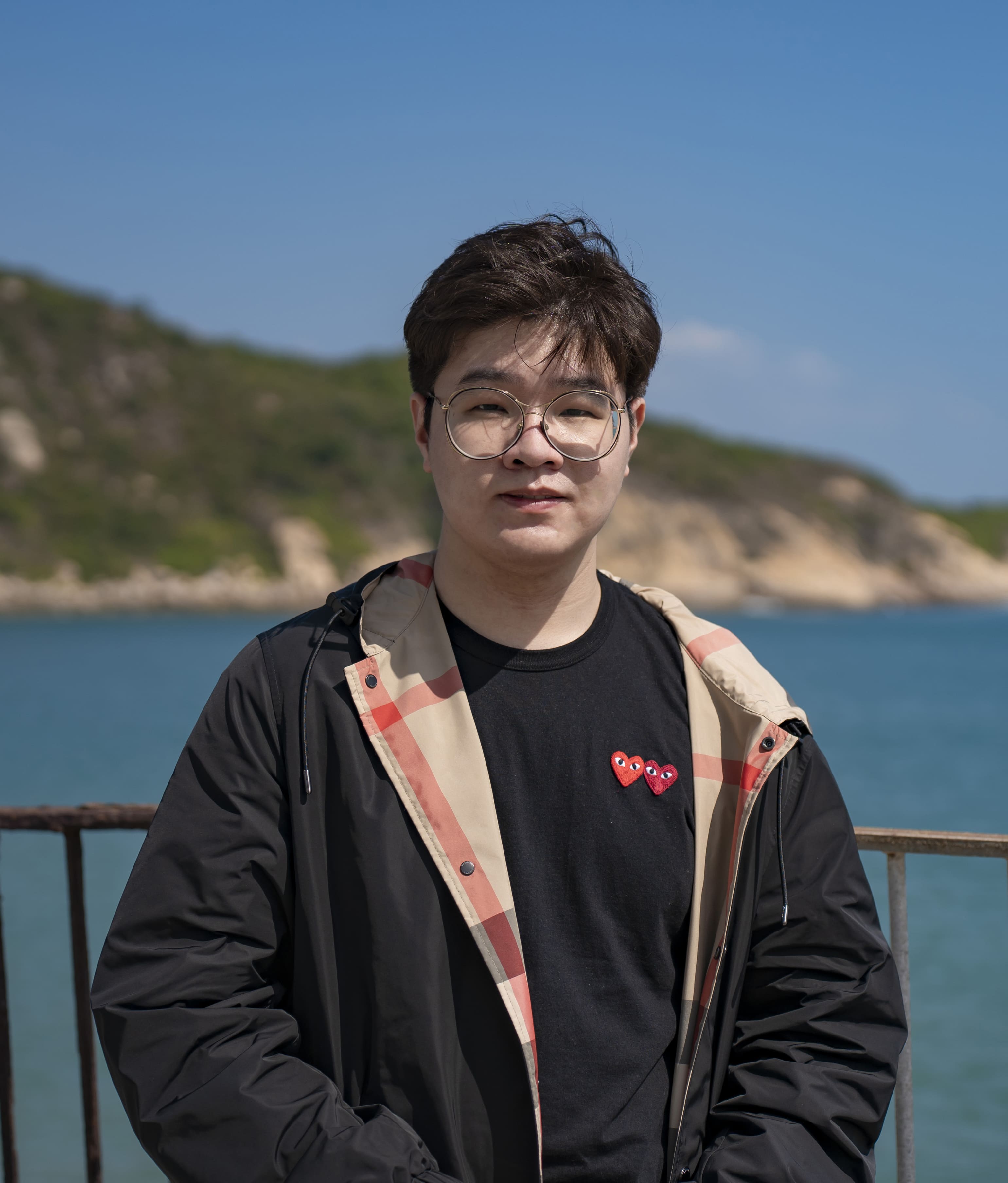}}]{Jiatong Li} is currently a PhD student of the Department of Computing (COMP), The Hong Kong Polytechnic University (funded by HKPFS). Before joining the PolyU, he received my Master's degree of Information Technology (with Distinction) from the University of Melbourne, under the supervision of Dr. Lea Frermann. In 2021, he got his bachelor's degree in Information Security from Shanghai Jiao Tong University. His interest lies in Natural Language Processing, Drug Discovery, and Recommender Systems. He has published innovative works in top-tier conferences such as IJCAI and ACL. For more information, please visit https://phenixace.github.io/.

\end{IEEEbiography}

\vspace{-24pt}
{\begin{IEEEbiography}[{\includegraphics[width=1in,height=1.25in,clip,keepaspectratio]{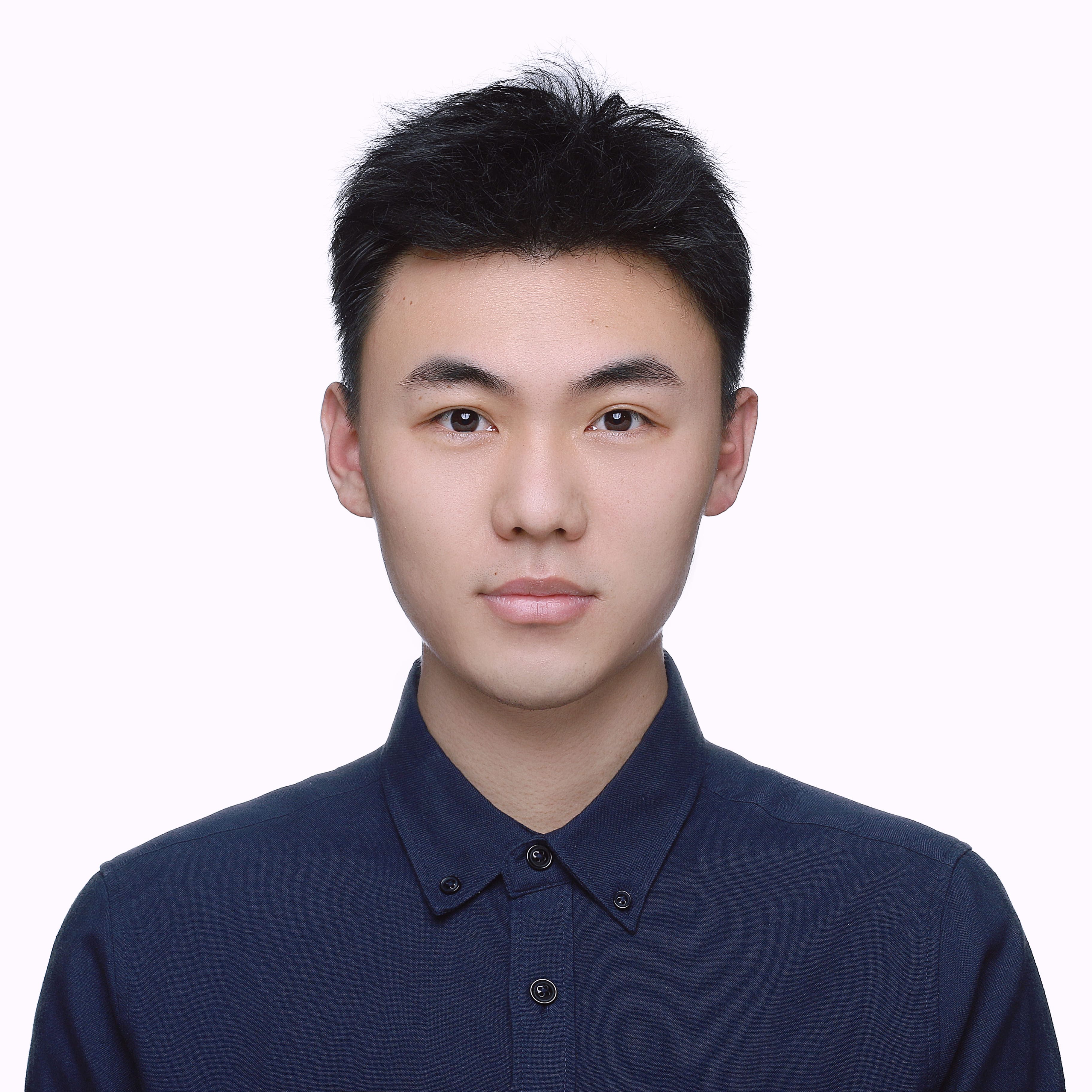}}]{Yunqing Liu} is currently a PhD student of the Department of Computing (COMP), Hong Kong Polytechnic University (PolyU), under the supervision of Dr. Wenqi Fan. Before joining the PolyU, he received his Master’s degree in Computer Science from the University of Edinburgh (M.Sc. in Computer Science), under the supervision of Dr. Elizabeth Polgreen. In 2020, he got his bachelor’s degrees from Wuhan University (B.Sc. in Chemistry and B.Eng. in Computer Science and Technology). His research interest includes Drug Discovery, Graph Neural Networks, and Natural Language Processing. He has published innovative works in top-tier conferences and journals such as IJCAI, EACL, EurJOC and Organic Letters. For more information, please visit https://liuyunqing.github.io/.

\end{IEEEbiography}

{\begin{IEEEbiography}[{\includegraphics[width=1in,height=1.25in,clip,keepaspectratio]{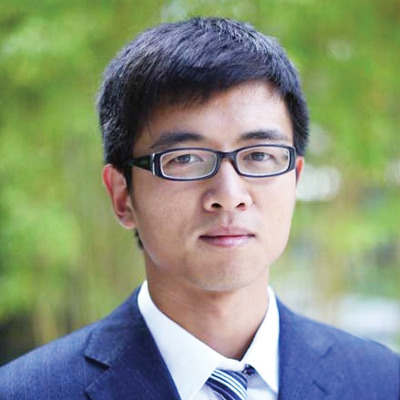}}]{Xiaowei Mei} received his PhD in Information Systems and Operations Management from the University of Florida. His current research aims to extend standard economic models of information systems in two directions: differentiating various forms of social contagion or peer effects in online and offline networks using empirical methods and big data analytic skills; and designing optimal market mechanisms in information systems using game theory, statistics and simulations methods. His work has been accepted by leading journals such as the Journal of Management Information Systems.
\end{IEEEbiography}

{\begin{IEEEbiography}[{\includegraphics[width=1in,height=1.25in,clip,keepaspectratio]{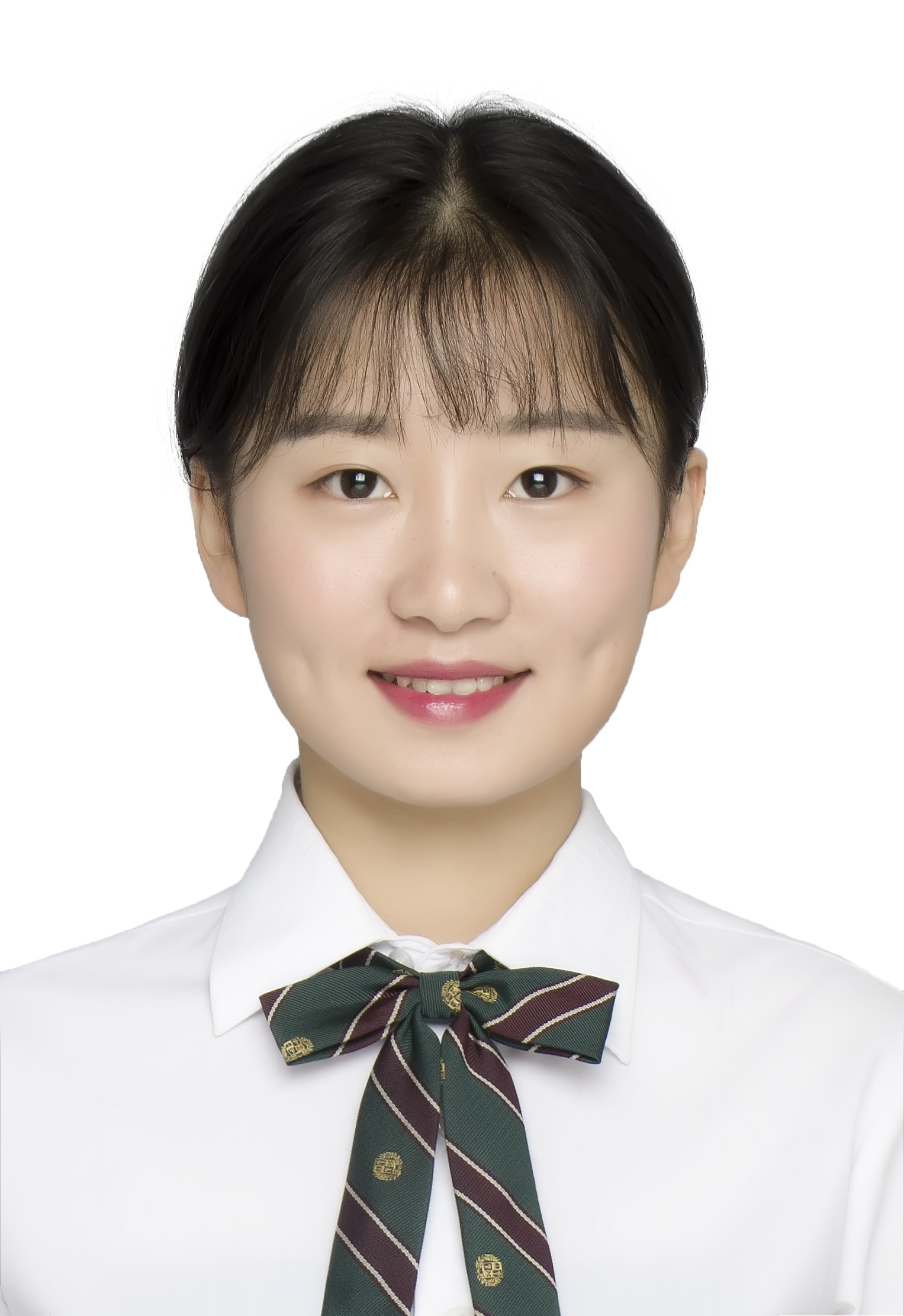}}]{Yiqi Wang} is an assistant professor at College of Computer, National University of Defense Technology (NUDT). She is currently working on graph neural networks including fundamental algorithms, robustness and their applications. She has published innovative works in top-tier conferences such as ICML, KDD, WWW, EMNLP, WSDM, and AAAI. She serves as top-tier conference program committee members (e.g., WWW, AAAI, IJCAI, CIKM, and WSDM) and journal reviewers (e.g., TIST, TKDD,  TKDE and TOIS). She also serves as the leading tutor of tutorials in top-tier conferences (e.g., KDD 2020, AAAI2021, SDM 2021, KDD 2021 and ICAPS 2021).  
\end{IEEEbiography}

{\begin{IEEEbiography}[{\includegraphics[width=1in,height=1.25in,clip,keepaspectratio]{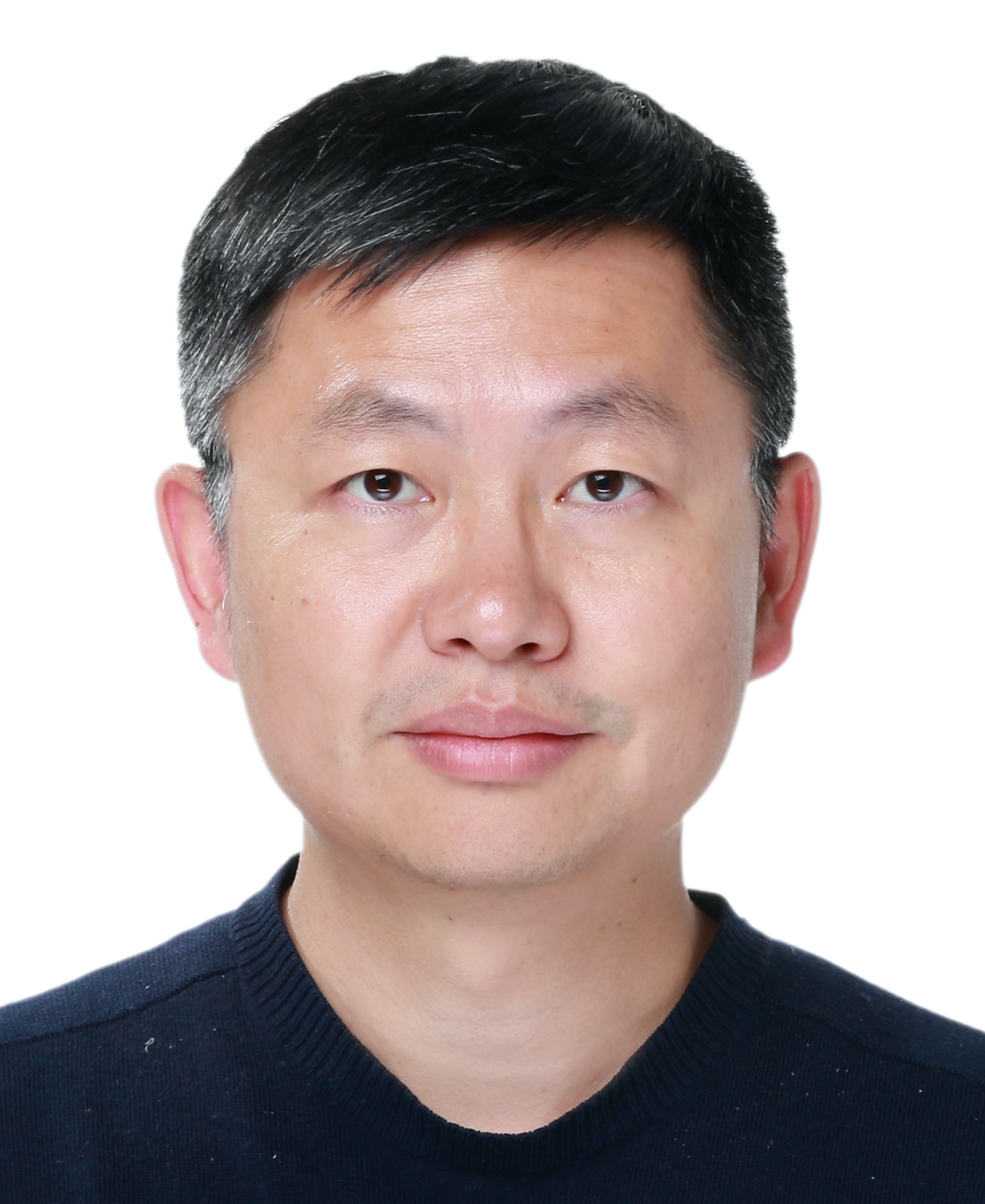}}]{Zhen Wen} is a Sr. Applied Science Manager at Amazon Prime Video, leading science efforts in video search, recommendation and promotions.  chief scientist of Tencent news feeds product, serving more than 200 million users each day. Dr. Wen directs a team of AI scientists and engineers aiming at deep content understanding, to source and push content users find most relevant and interesting.  Prior to his current role, he directed a team of AI scientists and engineers aiming at deep content understanding for short-form video recommendation at Tencent. He also held various science and technology roles at Alibaba Cloud, Google and IBM Research. Dr. Wen received PhD from University of Illinois at Urbana-Champaign. His work received best paper awards at International Conference On Information Systems and ACM Conference on Intelligent User Interfaces. Dr. Wen also received multiple Tencent Outstanding RD Award, IBM Outstanding Innovation Award, IBM Research Accomplishment Award, IBM invention achievement award. Dr. Wen served as an Associate Editor of IEEE Transactions on Multimedia.
\end{IEEEbiography}

\vspace{-24pt}
{\begin{IEEEbiography}[{\includegraphics[width=1in,height=1.25in,clip,keepaspectratio]{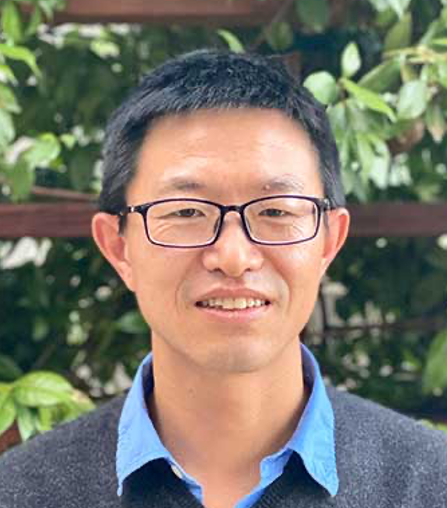}}]{Fei Wang} is head of personalization science at Amazon Prime Video responsible for improving user’s experience and engagement by developing a deep understanding of our customers and providing relevant, personalized and timely recommendations. Previously, he was a senior director with Visa Research leading a group of AI researchers to work on projects ranging from personalized restaurant recommendations, and fraud reduction, to credit risk prevention. With 50+ patents and 50+ research articles, he is also known for research on financial data mining, mobile healthcare, social computing and multimodal information retrieval.  He has received a number of best paper awards from conferences like RecSys, Multimedia Information Retrieval and Computers in Cardiology.
\end{IEEEbiography}

\vspace{-24pt}
\begin{IEEEbiography}[{\includegraphics[width=1in,height=1.25in,clip,keepaspectratio]{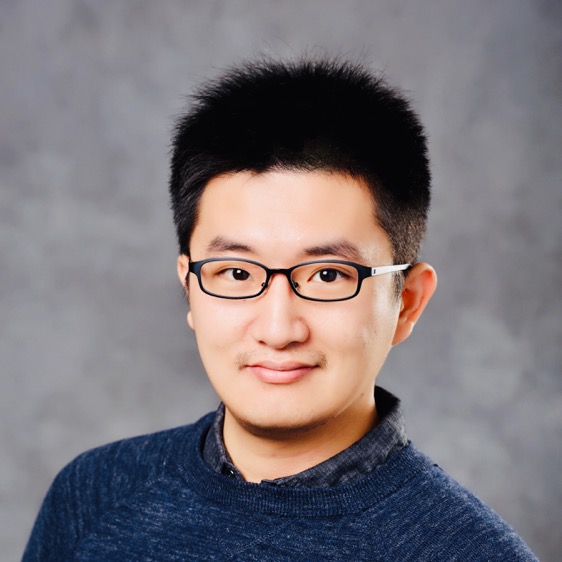}}]{Xiangyu Zhao} is an assistant professor of the school of data science at City University of Hong Kong (CityU). Before CityU, he completed his Ph.D. at Michigan State University. 
His current research interests include data mining and machine learning, especially on Reinforcement Learning and its applications in Information Retrieval.
He has published papers in top conferences (e.g., KDD, WWW, AAAI, SIGIR, ICDE, CIKM, ICDM, WSDM, RecSys, ICLR) and journals (e.g., TOIS, SIGKDD, SIGWeb, EPL, APS). His research received ICDM'21 Best-ranked Papers, Global Top 100 Chinese New Stars in AI, CCF-Tencent Open Fund, Criteo Research Award, and Bytedance Research Award.
He serves as top data science conference (senior) program committee members and session chairs (e.g., KDD, AAAI, IJCAI, ICML, ICLR, CIKM), and journal reviewers (e.g., TKDE, TKDD, TOIS, CSUR). 
He is the organizer of DRL4KDD@KDD’19, DRL4IR@SIGIR’20, 2nd DRL4KD@WWW’21, 2nd DRL4IR@SIGIR’21, and a lead tutor at WWW’21 and IJCAI’21. 
More information about him can be found at https://zhaoxyai.github.io/.
\end{IEEEbiography}

\begin{IEEEbiography}
[{\includegraphics[width=1in,height=1.25in,clip,keepaspectratio]{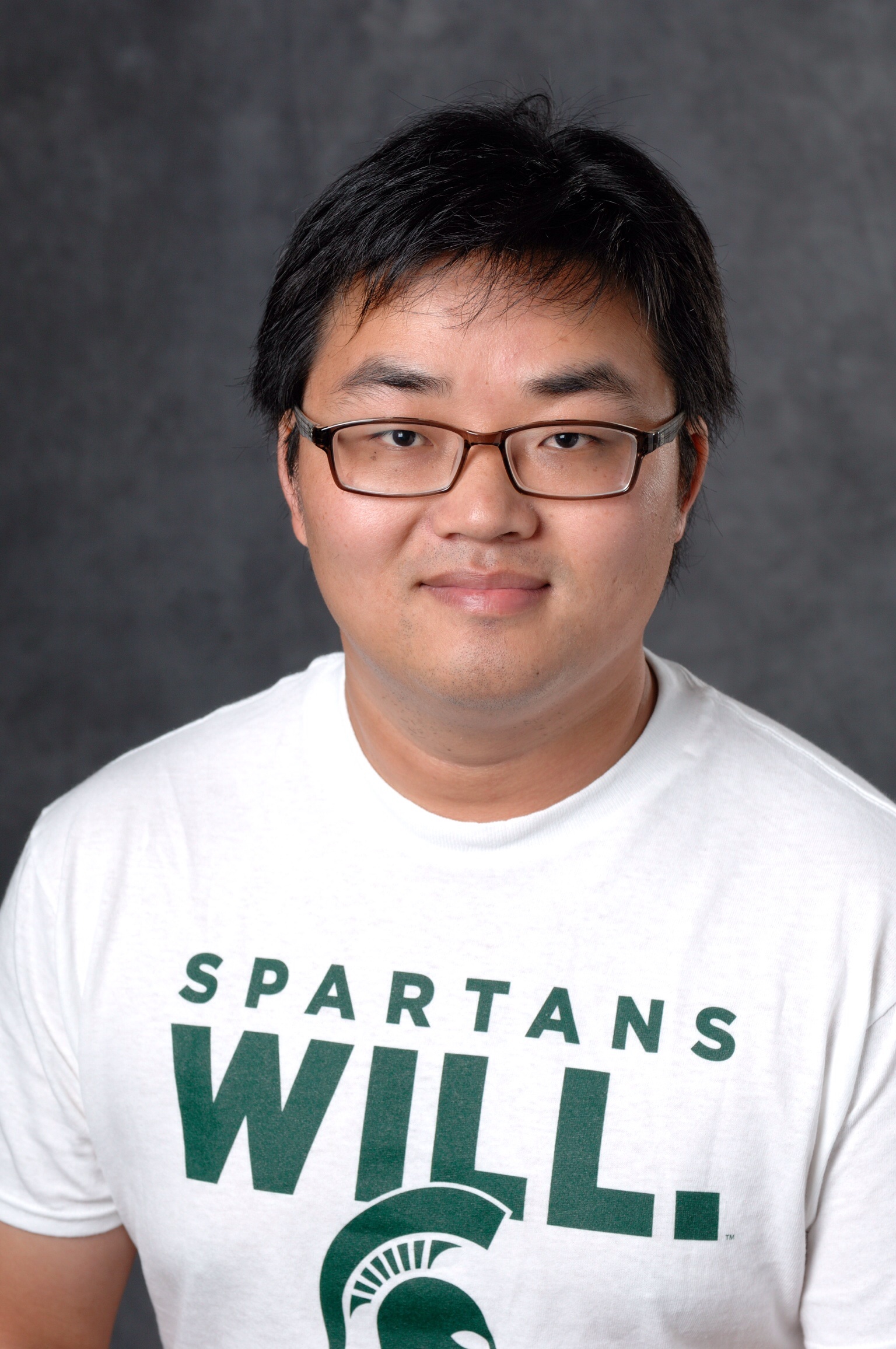}}]{Jiliang Tang} is a University Foundation Professor in the computer science and engineering department at Michigan State University since 2022. He was an associate professor (2021-2022) and an assistant professor (2016-2021) in the same department. Before that, he was a research scientist in Yahoo Research and got his PhD from Arizona State University in 2015 under Dr. Huan Liu. His research interests include graph machine learning, trustworthy AI and their applications in education and biology. He was the recipient of various awards including 2022 AI's 10 to Watch, 2022 IAPR J. K. AGGARWAL Award, 2022 SIAM/IBM Early Career Research Award, 2021 IEEE ICDM Tao Li Award, 2021 IEEE Big Data Security Junior Research Award, 2020 ACM SIGKDD Rising Star Award, 2020 Distinguished Withrow Research Award, 2019 NSF Career Award, and 8 best paper awards (or runner-ups). His dissertation won the 2015 KDD Best Dissertation runner up and Dean's Dissertation Award. He serves as conference organizers (e.g., KDD, SIGIR, WSDM and SDM) and journal editors (e.g., TKDD, TOIS and TKDE). He has published his research in highly ranked journals and top conference proceedings, which have received tens of thousands of citations with h-index 82 (Google Scholar) and extensive media coverage. More details about him can be found at https://www.cse.msu.edu/$\sim$tangjili/.
\end{IEEEbiography}


 \vspace{-24pt}
\begin{IEEEbiography}
[{\includegraphics[width=1in,height=1.25in,clip,keepaspectratio]{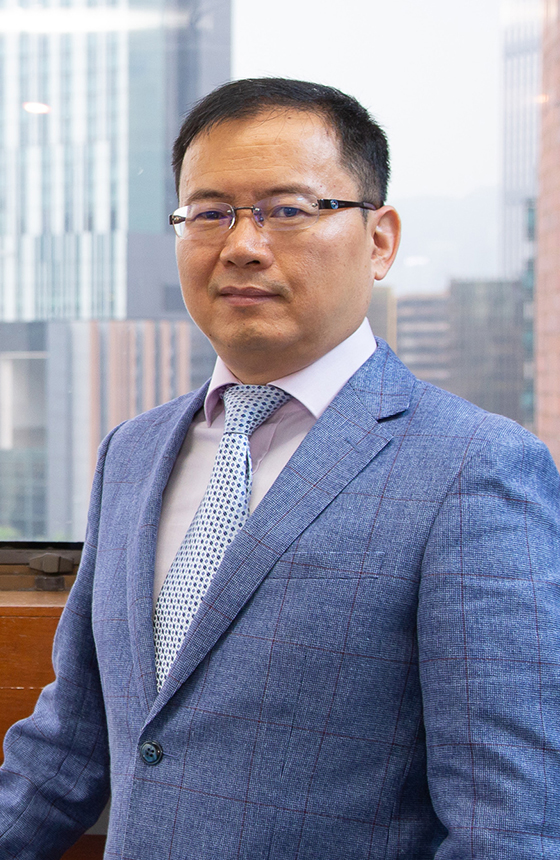}}]{Qing Li}
received the B.Eng. degree from Hunan University, Changsha, China, and the M.Sc. and Ph.D. degrees from the University of Southern California, Los Angeles, all in computer science.
He is currently a Chair Professor (Data Science) and the Head of the Department of Computing, the Hong Kong Polytechnic University. He is a Fellow of IEEE and IET, a member of ACM SIGMOD and IEEE Technical Committee on Data Engineering. 
His research interests include object modeling, multimedia databases, social media, and recommender systems. 
He has been actively involved in the research community by serving as an associate editor and reviewer for technical journals, and as an organizer/co-organizer of numerous international conferences. 
He is the chairperson of the Hong Kong Web Society, and also served/is serving as an executive committee (EXCO) member of IEEE-Hong Kong Computer Chapter and ACM Hong Kong Chapter. In addition, he serves as a councilor of the Database Society of Chinese Computer Federation (CCF), a member of the Big Data Expert Committee of CCF, and is a Steering Committee member of DASFAA, ER, ICWL, UMEDIA, and WISE Society. 
\end{IEEEbiography}
}

\end{document}